\documentclass[10pt, prd,twocolumn, nofootinbib,preprint,superscriptaddress]{revtex4}
\pdfoutput=1
\usepackage{amsmath,amssymb}
\usepackage{epsfig}
\usepackage{graphicx}
\usepackage[usenames,dvipsnames]{color}
\usepackage{subfigure}
\usepackage{slashed}
\usepackage[colorlinks,citecolor=blue]{hyperref}
\usepackage{color}

\begin{document}
	\title{Lepton Anomalous Magnetic Moment with Singlet-Doublet Fermion Dark Matter in Scotogenic $U(1)_{L_{\mu}-L_{\tau}}$ Model}

	\author{Debasish Borah}
	\email{dborah@iitg.ac.in}
	\affiliation{Department of Physics, Indian Institute of Technology Guwahati, Assam 781039, India}
	
	\author{Manoranjan Dutta}
	\email{ph18resch11007@iith.ac.in}
	\affiliation{Department of Physics, Indian Institute of Technology Hyderabad, Kandi, Sangareddy 502285, Telangana, India}
	
	\author{Satyabrata Mahapatra}
	\email{ph18resch11001@iith.ac.in}
	\affiliation{Department of Physics, Indian Institute of Technology Hyderabad, Kandi, Sangareddy 502285, Telangana, India}
	
	\author{Narendra Sahu}
	\email{nsahu@phy.iith.ac.in}
	\affiliation{Department of Physics, Indian Institute of Technology Hyderabad, Kandi, Sangareddy 502285, Telangana, India}
	
	\begin{abstract}
		We study an extension of the minimal gauged $L_{\mu}-L_{\tau}$ model  including three right handed singlet fermions and a 
		scalar doublet to explain the anomalous magnetic moments of muon and electron simultaneously. The presence of an in-built $Z_2$ symmetry under which 
		the right handed singlet fermions and $\eta$ are odd, gives rise to a stable dark matter candidate along with light neutrino mass in a scotogenic fashion.
		In spite of the possibility of having positive and negative contributions to  muon and electron $(g-2)$ respectively from vector boson and charged 
		scalar loops, the minimal scotogenic $L_{\mu}-L_{\tau}$ model can not explain both muon and electron $(g-2)$ simultaneously while being consistent with other experimental bounds. We then extend the model with a vector like lepton doublet which not only leads to a chirally enhanced negative contribution to electron $(g-2)$ but also leads to the popular singlet-doublet fermion dark matter scenario. With this extension, the model can explain both electron and muon $(g-2)$ while being consistent with neutrino mass, dark matter and other direct search bounds. The model remains predictive at high energy experiments like collider as well as low energy experiments looking for charged lepton flavour violation, dark photon searches, in addition to future $(g-2)$ measurements.
	\end{abstract}
	
	\maketitle
	\noindent
	\section{Introduction}
	The muon anomalous magnetic moment, $a_\mu$ = $(g - 2)_\mu/2$ has been measured recently by the E989
	experiment at Fermi lab showing a discrepancy with respect to the theoretical prediction of the Standard
	Model (SM) \cite{Abi:2021gix}
	\begin{eqnarray}
		a^{\rm FNAL}_\mu = 116 592 040(54) \times 10^{-11}\\
		a^{\rm SM}_\mu = 116 591 810(43) \times 10^{-11}
	\end{eqnarray}
	which, when combined with the previous Brookhaven determination of
	\begin{equation}
		a^{\rm BNL}_\mu = 116 592 089(63) \times 10^{-11}
	\end{equation}
	leads to a 4.2 $\sigma$ observed excess of
	$\Delta a_\mu = 251(59) \times 10^{-11}$ \footnote{The is however, in contrast with the latest lattice results \cite{Borsanyi:2020mff} predicting a larger value of muon $(g-2)$ keeping it closer to experimental value. Measured value of muon $(g-2)$ is also in tension with  global electroweak fits from $e^+ e^-$ to hadron data \cite{Crivellin:2020zul, Colangelo:2020lcg, Keshavarzi:2020bfy}.}. The status of the SM calculation of muon magnetic moment has been updated recently in \cite{Aoyama:2020ynm}. While the muon anomalous magnetic moment has been known for a long time, the recent Fermilab measurement has led to several recent works. Review of such theoretical explanations for muon $(g-2)$ can be found in \cite{Jegerlehner:2009ry, Lindner:2016bgg, Athron:2021iuf}. Gauged lepton flavour models like $U(1)_{L_{\mu}-L_{\tau}}$ provide a natural origin of muon $(g-2)$ in a very minimal setup while also addressing the question of the origin of light neutrino mass and mixing \cite{Zyla:2020zbs} simultaneously. Recent studies on this model related to muon $(g-2)$ may be found in \cite{Borah:2020jzi, Zu:2021odn, Amaral:2021rzw, Zhou:2021vnf, Borah:2021jzu, Borah:2021mri, Holst:2021lzm, Singirala:2021gok, Hapitas:2021ilr,Kang:2021jmi}.
	
	Interestingly, similar anomaly in electron magnetic moment has also been reported from a recent precision measurement of the fine structure constant using Cesium atoms \cite{Parker:2018vye}. Using the precise measured value of fine structure constants lead to a different SM predicted value for electron anomalous magnetic moment $a_e=(g-2)_e/2$. Comparing this with the existing experimental value of $a_e$ leads to a discrepancy
	\begin{equation}
		\Delta a_e =a^{\rm exp}_e-a^{\rm SM}_e= (-87 \pm 36) \times 10^{-14}
	\end{equation}
	at $2.4\sigma$ statistical significance\footnote{Interestingly, a more recent measurement of the fine structure constant using Rubidium atoms have led to a milder $(g-2)_e$ anomaly at $1.6\sigma$ statistical significance, but in the opposite direction \cite{Morel:2020dww}. Here we consider negative $(g-2)_e$ only}.  Several works have attempted to find a common explanation of electron and muon $(g-2)$ within well motivated particle physics frameworks \cite{Davoudiasl:2018fbb, Crivellin:2018qmi, Liu:2018xkx, Han:2018znu, Endo:2019bcj, Abdullah:2019ofw, Bauer:2019gfk, Badziak:2019gaf, CarcamoHernandez:2019ydc, Hiller:2019mou, Cornella:2019uxs, Endo:2020mev, Jana:2020pxx, Calibbi:2020emz, Yang:2020bmh, Chen:2020jvl, Hati:2020fzp, Dutta:2020scq, Botella:2020xzf, Chen:2020tfr, Dorsner:2020aaz, Arbelaez:2020rbq, Jana:2020joi, Chun:2020uzw, Li:2020dbg, DelleRose:2020oaa, Hernandez:2021tii, Bodas:2021fsy, Cao:2021lmj, Han:2021gfu, Escribano:2021css, De:2021crr}.
	
	Here we consider the popular and minimal model based on the gauged $L_{\mu}-L_{\tau}$ symmetry which is anomaly free \cite{He:1990pn, He:1991qd}.  Apart from the SM fermion content, the minimal version of this model has three heavy right handed neutrinos (RHN) leading to type I seesaw origin of light neutrino masses \cite{Minkowski:1977sc, Mohapatra:1979ia, Yanagida:1979as, GellMann:1980vs, Glashow:1979nm, Schechter:1980gr,Patra:2016shz}. We also consider an additional scalar doublet $\eta$ and an in-built $Z_2$ symmetry under which RHNs and $\eta$ are odd while SM fields are even. The unbroken $Z_2$ symmetry guarantees a stable dark matter (DM) candidate while light neutrino masses arise at one-loop in scotogenic fashion \cite{Ma:2006km}. A $L_{\mu}-L_{\tau}$ extension of the scotogenic model was discussed earlier in the context of DM and muon $(g-2)$ in \cite{Baek:2015fea}. We consider all possible contributions to $(g-2)_{e, \mu}$ in this model and show that it is not possible to satisfy them simultaneously while being consistent with phenomenological requirements like neutrino mass, lepton flavour violation (LFV) and direct search bounds. We then consider an extension of the minimal scotogenic $L_{\mu}-L_{\tau}$ model with a $Z_2$-odd vector like lepton doublet with vanishing $U(1)_{L_{\mu}-L_{\tau}}$ charge. While the minimal scotogenic extension was discussed in \cite{Baek:2015fea} mentioned above, the vector like lepton extension was recently discussed in \cite{Kang:2021jmi} from muon $(g-2)$ point of view. Here we extend this idea to include electron $(g-2)$ along with detailed study of DM and related phenomenology. Due to the possibility of vector like lepton doublet coupling with one of the RHNs, it gives rise to a singlet-doublet fermion DM scenario studied extensively in the literature \cite{Mahbubani:2005pt,DEramo:2007anh,Enberg:2007rp,Cohen:2011ec,Cheung:2013dua, Restrepo:2015ura,Calibbi:2015nha,Cynolter:2015sua, Bhattacharya:2015qpa,Bhattacharya:2017sml, Bhattacharya:2018fus,Bhattacharya:2018cgx,DuttaBanik:2018emv,Barman:2019tuo,Bhattacharya:2016rqj, Calibbi:2018fqf, Barman:2019aku, Dutta:2020xwn,Dutta:2021uxd}. We show that the inclusion of this vector like lepton doublet which couples only to electrons and one of the RHNs due to vanishing $U(1)_{L_{\mu}-L_{\tau}}$ charge, the observed $(g-2)_e$ can be generated due to chiral enhancement in the one-loop diagram mediated by charged scalar and fermions. On the other hand a positive muon $(g-2)$ can be generated due to light $U(1)_{L_{\mu}-L_{\tau}}$ gauge boson mediated one-loop diagram. We then discuss the relevant DM phenomenology and possibility of observable lepton flavour violation as well as some collider signatures. 
	

	This paper is organized as follows. In section \ref{sec2}, we discuss the minimal scotogenic $U(1)_{L_{\mu}-L_{\tau}}$ model and show that it is not possible to explain both electron and muon $(g-2)$ simultaneously. In section \ref{sec3}, we consider the extension of minimal model by a vector like lepton doublet and show its success in explaining the lepton $(g-2)$ data. We then discuss singlet-doublet DM phenomenology in section \ref{sec4} followed by brief discussion on collider phenomenology in section \ref{collider}. We finally conclude in section \ref{sec6}.

	\begin{table}[h!]
		\small
		\begin{center}
			\begin{tabular}{||@{\hspace{0cm}}c@{\hspace{0cm}}|@{\hspace{0cm}}c@{\hspace{0cm}}|@{\hspace{0cm}}c@{\hspace{0cm}}|@{\hspace{0cm}}c@{\hspace{0cm}}||}
				\hline
				\hline
				\begin{tabular}{c}
					{\bf ~~~~ Gauge~~~~}\\
					{\bf ~~~~Group~~~~}\\ 
					\hline
					
					$SU(2)_{L}$\\ 
					\hline
					$U(1)_{Y}$\\ 
					\hline
					$U(1)_{L_\mu-L_\tau}$\\ 
					\hline
					$Z_2$\\ 
				\end{tabular}
				&
				&
				\begin{tabular}{c|c|c}
					\multicolumn{3}{c}{\bf Fermion Fields}\\
					\hline
					~~~$N_e$~~~& ~~~$N_{\mu}$~~~ & ~~~$N_{\tau}$~~~ \\
					\hline
					$1$&$1$&$1$\\
					\hline
					$0$&$0$&$0$\\
					\hline
					$0$&$1$&$-1$\\
					\hline
					$-1$&$-1$&$-1$\\
				\end{tabular}
				&
				\begin{tabular}{c|c|c}
					\multicolumn{3}{c}{\bf Scalar Field}\\
					\hline
					~~~$\Phi_{1}$~~~& ~~~$\Phi_2$~~~ & ~~~$\eta$~~~\\
					\hline
					$1$&$1$&$2$\\
					\hline
					$0$&$0$&$\frac{1}{2}$\\
					\hline
					$1$&$2$&$0$\\
					\hline
					$+1$&$+1$&$-1$\\
				\end{tabular}\\
				\hline
				\hline
			\end{tabular}
			\caption{New Particles and their
				gauge charges in minimal scotogenic $U(1)_{L_{\mu}-L_{\tau}}$ model.}
			\label{tab1}
		\end{center}    
	\end{table}

	\noindent
	\section{The Minimal Model}
	\label{sec2}
	The SM fermion content with their gauge charges under $SU(3)_c \otimes SU(2)_L \otimes U(1)_Y \otimes U(1)_{L_{\mu}-L_{\tau}}$ gauge symmetry are denoted as follows.
	
	$$ q_L=\begin{pmatrix}u_{L}\\
		d_{L}\end{pmatrix} \sim (3, 2, \frac{1}{6}, 0), \; u_R (d_R) \sim (3, 1, \frac{2}{3} (-\frac{1}{3}), 0)$$
	$$L_e=\begin{pmatrix}\nu_{e}\\
		e_{L}\end{pmatrix} \sim (1, 2, -\frac{1}{2}, 0), \; e_R \sim (1, 1, -1, 0) $$
	$$L_{\mu}=\begin{pmatrix}\nu_{\mu}\\
		\mu_{L}\end{pmatrix} \sim (1, 2, -\frac{1}{2}, 1), \; \mu_R \sim (1, 1, -1, 1) $$
	$$L_{\tau}=\begin{pmatrix}\nu_{\tau}\\
		\tau_{L}\end{pmatrix} \sim (1, 2, -\frac{1}{2}, -1), \;  \tau_R \sim (1, 1, -1, -1)$$ 
	
	The new field content apart from the SM ones are shown in table \ref{tab1}. The SM fields are even under the $Z_2$ symmetry and only the second and third generations of leptons are charged under the $L_{\mu}-L_{\tau}$ gauge symmetry. The relevant Lagrangian can be written as
	\begin{widetext}
		\begin{align}
			\mathcal{L} & \supseteq \overline{N_{\mu}} i \gamma^\mu \mathfrak{D}_\mu N_{\mu} - M_{\mu \tau} N_{\mu} N_{\tau} +\overline{N_{\tau}} i \gamma^\mu \mathfrak{D}_\mu N_{\tau} - \frac{M_{ee}}{2} N_e N_e -Y_{e\mu} \Phi^{\dagger}_1 N_e N_\mu -Y_{e\tau} \Phi_1 N_e N_\tau -Y_{\mu} \Phi^{\dagger}_2 N_\mu N_\mu \nonumber \\&-Y_{De} \bar{L_e} \tilde{\eta} N_e-Y_{D\mu} \bar{L_\mu} \tilde{\eta} N_\mu-Y_{D \tau} \bar{L_\tau} \tilde{\eta} N_\tau-Y_{\tau} \Phi_2 N_\tau N_\tau - Y_{le} \overline{L_e} H e_R - Y_{l\mu} \overline{L_\mu} H \mu_R
			- Y_{l\tau} \overline{L_\tau} H \tau_R +{\rm h.c.}
			\label{eq:yuklag} \end{align}
	\end{widetext}
	where $H$ is the SM Higgs doublet and the covariant derivative $\mathfrak{D}_\mu$ is given as 
	\begin{equation}
		\begin{aligned}
			\mathfrak{D}_\mu&=\partial_\mu - ig_{\mu \tau} Y_{\mu \tau} (Z_{\mu \tau})_\mu\,.
		\end{aligned}
	\end{equation}
	Also, in the above Lagrangian, $\widetilde{\eta}=i\tau_2 \eta^{*}$. The new gauge kinetic terms that appear in the Lagrangian constitute of,
	\begin{equation}
		\mathcal{L}_{\rm Gauge} = -\frac{1}{4} (Z_{\mu \tau})_{\mu\nu}Z^{\mu \nu}_{\mu \tau} - \frac{\epsilon}{2} (Z_{\mu \tau})_{\mu\nu} B^{\mu \nu};
	\end{equation}
	where, $\epsilon$ parametrises the kinetic mixing between the $U(1)_{L_\mu - L_\tau}$ and $U(1)_Y$ gauge sectors.
	
	The Lagrangian of scalar sector is given by:
	\begin{equation}
		\mathcal{L}_{scalar}= |\mathcal{D}_\mu{H}|^2+|\mathcal{D}_\mu{\eta}|^2+|\mathfrak{D}_\mu{\Phi}_{i}|^2 - V(H,\Phi_{i}, \eta)
	\end{equation}
	where $i=1,2$. The covariant derivatives $\mathcal{D}_\mu$ is given as follows:
	\begin{equation}
		\begin{aligned}
			\mathcal{D}_\mu&=\partial_\mu  - i\frac{g}{2}\tau.W_\mu - ig'\frac{Y}{2}B_\mu
		\end{aligned}
	\end{equation}
	
	The scalar potential is given by
	\begin{equation}
		\begin{aligned}
			V(H, \Phi_{i},\eta)&= -{\mu^2_H} \left(H^\dagger H \right) + \lambda_H \left(H^\dagger H \right)^2 -{\mu^2_{\Phi_i}} ({\Phi_{i}}^\dagger \Phi_{i} ) \nonumber \\
			&+ \lambda_{\Phi_i} ({\Phi_{i}}^\dagger \Phi_{i} )^2 \nonumber  + \lambda_{H \Phi_i} (H^\dagger H)({\Phi_{i}}^\dagger \Phi_{i} )\,\nonumber\\&+m^2_{\eta}(\eta^\dagger \eta)+\lambda_2 (\eta^\dagger \eta)^2+\lambda_{3} (\eta^{\dagger} \eta) (H^{\dagger} H)\nonumber \\& + \lambda_4 (\eta^{\dagger} H) (H^{\dagger} \eta) 
			+ \dfrac{\lambda_{5}}{2} [ (H^{\dagger}\eta)^2+ (\eta^{\dagger}H)^2]\nonumber \\&
		+\lambda_{\eta \Phi_i} (\eta^\dagger \eta)(\Phi_{i}^\dagger \Phi_{i}) + \lambda_{\Phi_1\Phi_2}(\Phi_1^\dagger \Phi_1)(\Phi_2^\dagger \Phi_2) \nonumber \\& + [\mu_{12}\Phi^2_1  \Phi^\dagger_2+{\rm h.c.}],
		\end{aligned}
	\end{equation}
	where $i=1,2$ denotes two singlet scalars $\Phi_{1,2}$. While the neutral component of the Higgs doublet $H$ breaks the electroweak gauge symmetry, the singlets $\Phi_{1,2}$ break $L_{\mu}-L_{\tau}$ gauge symmetry after acquiring non-zero vacuum expectation values (VEV). Denoting the VEVs of singlets $\Phi_{1,2}$ as $v_{1,2}$, the new gauge boson mass can be found to be $M_{Z_{\mu \tau}}=g_{\mu \tau} \sqrt{2(v^2_1+4v^2_2)}$ with $g_{\mu \tau}$ being the $L_{\mu}-L_{\tau}$ gauge coupling. Clearly the model predicts diagonal charged lepton mass matrix $M_\ell$ and diagonal Dirac Yukawa of neutrinos. Thus, the non-trivial neutrino mixing will arise from the structure of right handed neutrino mass matrix $M_R$ only which is generated by the chosen scalar singlet fields. The right handed neutrino mass matrix, Dirac neutrino Yukawa and charged lepton mass matrix are given by
	\begin{align}
		M_R =\begin{pmatrix}
			M_{ee}      &  Y_{e\mu} v_1
			& Y_{e\tau} v_ 1 \\
			Y_{e\mu} v_ 1      &  Y_{\mu}
			v_2     & M_{\mu \tau}  \\
			Y_{e\tau} v_1     &  M_{\mu \tau}    &
			Y_{\tau} v_2  
		\end{pmatrix}\, \nonumber \\
		Y_D =\begin{pmatrix}
			Y_{De}       &  0    & 0  \\
			0     &  Y_{D\mu}    & 0  \\
			0     &  0    & Y_{D \tau}  
		\end{pmatrix},\,              M_\ell= \frac{1}{\sqrt{2}}\begin{pmatrix}
			Y_{e} v & 0 & 0\\
			0 & Y_{\mu}v & 0 \\
			0 & 0 & Y_{\tau} v
		\end{pmatrix}
		\label{mass_mat}
	\end{align}
	Here $v/\sqrt{2}$ is the VEV of neutral component of SM Higgs doublet $H$.
	
	The one-loop neutrino mass can be written as \cite{Ma:2006km, Merle:2015ica}
	\begin{align}
		(M_{\nu})_{ij} \ & = \ \sum_k \frac{h_{ik}h_{jk} M_{k}}{32 \pi^2} \left[L_k(m^2_{\eta_R})-L_k(m^2_{\eta_I})\right] \, ,
		\label{numass1}
	\end{align}
	where 
	$M_k$ is the mass eigenvalue of the RHN mass eigenstate $N_k$ in the internal line and the indices $i, j = 1,2,3$ run over the three neutrino generations. The Yukawa couplings appearing in the neutrino mass formula above are derived from the corresponding Dirac Yukawa couplings in Lagrangian \eqref{eq:yuklag} by going to the diagonal basis of right handed neutrinos after spontaneous symmetry breaking. The loop function $L_k(m^2)$ in neutrino mass formula \eqref{numass1} is defined as 
	\begin{align}
		L_k(m^2) \ = \ \frac{m^2}{m^2-M^2_k} \: \text{ln} \frac{m^2}{M^2_k} \, .
		\label{eq:Lk}
	\end{align}
	The difference in masses for neutral scalar and pseudo-scalar components of $\eta$, crucial to generate non-zero neutrino mass is $m^2_{\eta_R}-m^2_{\eta_I}=\lambda_5 v^2$.  We first diagonalise $M_R$ and consider the physical basis of right handed neutrinos $(N_1, N_2, N_3)$ with appropriate interactions. We also use Casas-Ibarra (CI) parametrisation \cite{Casas:2001sr} extended to radiative seesaw model \cite{Toma:2013zsa} which allows us to write the Yukawa coupling matrix satisfying the neutrino data as
	\begin{align}
		h_{\alpha i} \ = \ \left(U D_\nu^{1/2} R^{\dagger} \Lambda^{1/2} \right)_{\alpha i} \, ,
		\label{eq:Yuk}
	\end{align}
	where $R$ is an arbitrary complex orthogonal matrix satisfying $RR^{T}=\mathrm{I}$. Here $D_\nu \ =  \textrm{diag}(m_1,m_2,m_3)$ is the diagonal light neutrino mass matrix and the diagonal matrix $\Lambda$ has elements given by
	\begin{align}
		\Lambda_k \ & = \ \frac{2\pi^2}{\lambda_5}\zeta_k\frac{2M_k}{v^2} \, , \\
		\textrm {and}\quad \zeta_k & \ = \  \left(\frac{M_{k}^2}{8(m_{\eta_R}^2-m_{\eta_I}^2)}\left[L_k(m_{\eta_R}^2)-L_k(m_{\eta_I}^2) \right]\right)^{-1} \, . \label{eq:zeta}
	\end{align}
	

	\subsection{Anomalous Magnetic Moment} The magnetic moment of muon is defined as
	\begin{equation}\label{anomaly}
		\overrightarrow{\mu_\mu}= g_\mu \left (\frac{q}{2m} \right)
		\overrightarrow{S}\,,
	\end{equation}
	where $g_\mu$ is the gyro-magnetic ratio and its value is $2$ for an elementary spin $\frac{1}{2}$ particle of mass $m$ and charge
	$q$. However, higher order radiative corrections can generate additional contributions to its magnetic moment and is parameterized as
	\begin{equation}
		a_\mu=\frac{1}{2} ( g_\mu - 2).
	\end{equation}

	\begin{figure}[h!]
		\centering
		\includegraphics[scale=0.45]{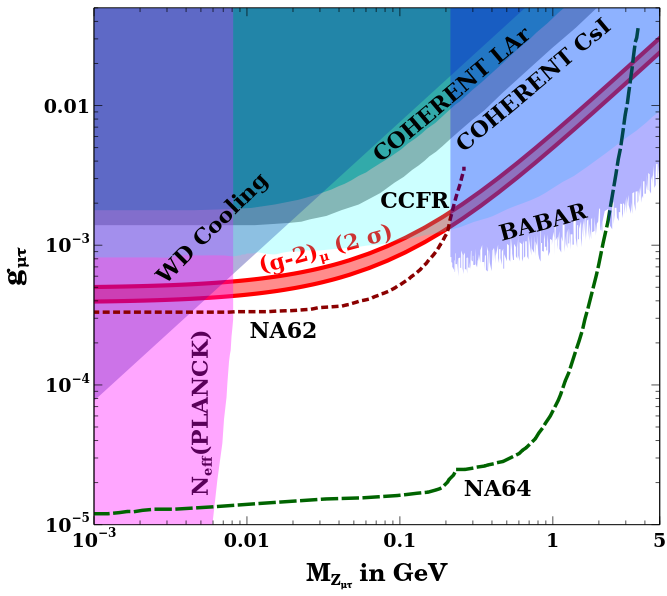}
		\caption{Parameter space satisfying $(g-2)_\mu$ in the plane of $g_{\mu\tau}-M_{Z_{\mu\tau}}$. See text for details related to other bounds imposed.}
		\label{fig:muong2}
	\end{figure}
	
	As mentioned earlier, the anomalous muon magnetic moment has been measured very precisely in the recent Fermilab experiment while it has also been predicted in the SM to a great accuracy. In the model under consideration in this work, the additional contribution to muon magnetic moment arises dominantly from one-loop diagram mediated by $L_{\mu}-L_{\tau}$ gauge boson $Z_{\mu \tau}$. The corresponding one-loop contribution is given by \cite{Brodsky:1967sr, Baek:2008nz}
	\begin{equation}
		\Delta a_{\mu} = \frac{\alpha'}{2\pi} \int^1_0 dx \frac{2m^2_{\mu} x^2 (1-x)}{x^2 m^2_{\mu}+(1-x)M^2_{Z_{\mu \tau}}} \approx \frac{\alpha'}{2\pi} \frac{2m^2_{\mu}}{3M^2_{Z_{\mu \tau}}}
	\end{equation}
	where $\alpha'=g^2_{\mu \tau}/(4\pi)$.
	\begin{figure}[h!]
		\centering
		\includegraphics[scale=0.3]{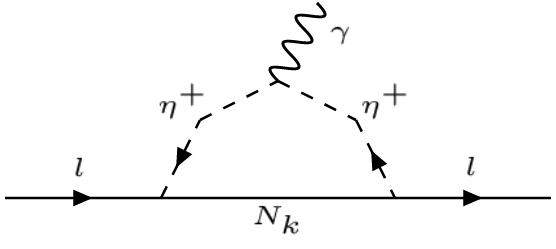}
		\caption{Negative contribution to $\Delta a_{l}$ coming from charged scalar $\eta^+$ and neutral fermion $N_k$ in the loop.}
		\label{fig:scotog2}
	\end{figure} 
	\begin{figure}[h!]
		\centering
		\includegraphics[scale=0.5]{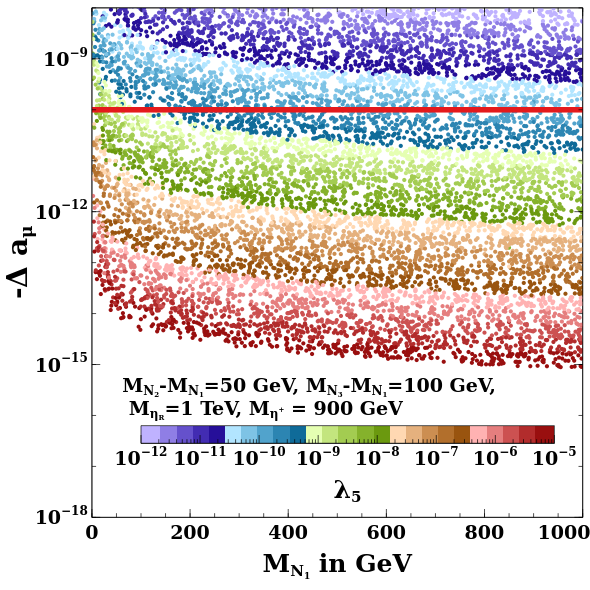}
		\caption{Negative contribution to $\Delta a_{\mu}$ (coming from charged scalar $\eta^+$ and neutral fermion $N_k$ in the loop) is shown as a function of $N_1$ mass. The colour bar shows the values of $\lambda_{5}$.}
		\label{fig:lam5amu}
	\end{figure}

	The parameter space satisfying muon $g-2$ in the plane of $g_{\mu \tau}$ versus $M_{Z_{\mu\tau}} $ is shown in Fig~\ref{fig:muong2}.  Several exclusion limits from different experiments namely, CCFR \cite{Altmannshofer:2014pba}, COHERENT \cite{Akimov:2017ade, Akimov:2020pdx},  BABAR \cite{TheBABAR:2016rlg} are also shown. The astrophysical bounds from cooling of white dwarf (WD) \cite{Bauer:2018onh, Kamada:2018zxi} excludes the upper left triangular region. Very light $Z'$ is ruled out from cosmological constraints on effective relativistic degrees of freedom \cite{Aghanim:2018eyx, Kamada:2018zxi, Ibe:2019gpv, Escudero:2019gzq}. This arises due to the late decay of such light gauge bosons into SM leptons, after standard neutrino decoupling temperatures thereby enhancing $N_{\rm eff}$. Future sensitivities of NA62 \cite{Krnjaic:2019rsv} and NA64 \cite{Gninenko:2014pea, Gninenko:2018tlp} experiments are also shown as dashed lines. In summary, the model has a small parameter space, currently allowed from all experimental bounds, which can explain muon $(g-2)$. This has also been noticed in earlier works on minimal $L_{\mu}-L_{\tau}$ model \cite{Borah:2020jzi, Zu:2021odn, Amaral:2021rzw, Zhou:2021vnf, Borah:2021jzu, Borah:2021mri, Holst:2021lzm, Singirala:2021gok, Hapitas:2021ilr}.
	
	However, note that in scotogenic version of $L_{\mu}-L_{\tau}$ model we can have another one-loop diagram mediated by charged component of scalar doublet $\eta$ and right handed neutrino $N_{k}$ contributing to $(g-2)$. Since both electron and muon couple to charged scalar via Yukawa couplings, we can have contributions to the anomalous magnetic moments of both electron and muon via this diagram, shown in Fig.~\ref{fig:scotog2}. It is given by \cite{Queiroz:2014zfa, Calibbi:2018rzv, Jana:2020joi}
	\begin{equation}
		\Delta a_l =\sum_k -\frac{m^2_l}{8\pi^2 M^2_{\eta^+}} \lvert h_{lk} \rvert^2 f(M^2_{k}/M^2_{\eta^+})
	\end{equation}
	where 
	\begin{equation}
		f(x)=\frac{1-6x+3x^2+2x^3-6x^2\log{x}}{12 (1-x)^4}
		\label{loop1}
	\end{equation}
	which gives rise to an overall negative contribution to $(g-2)_l$. While this works for electron $(g-2)$, for muon one needs to make sure that the positive contribution coming from vector boson loop dominates over the one from the charged scalar loop so that the total muon $(g-2)$ remains positive as suggested by experiments. 
	
	Also from Fig.~\ref{fig:muong2}, we can see that there is still a small parameter space left with $M_{Z_{\mu \tau}}$ in a range of $10-30$ MeV and $g_{\mu \tau}$ in $(5.5-7.5)\times 10^{-4}$ which is not constrained by CCFR and can accommodate a larger positive contribution to muon $(g-2)$ from vector boson loop. This positive contribution to $\Delta a_\mu$ can be as large as $5.5\times10^{-9}$ which is $2.4\times 10^{-9}$ larger than the current upper limit of $3.1\times 10^{-9}$. Hence, even if we get a negative contribution from the $\eta^+$ loop to muon $(g-2)$ which is of the order $\mathcal{O}(10^{-9})$ but smaller than $3.5\times 10^{-9}$, we can still get the correct value of $\Delta a_\mu$ as measured by the experiments. Thus, in general, observed muon $(g-2)$ can be explained by a combination of positive and negative contributions in scotogenic $L_{\mu}-L_{\tau}$ model.
	
	It is worth mentioning here that $\lambda_{5}$ plays a crucial role in the contribution of charged scalar loop into $(g-2)$ as it decides the strength of Yukawa couplings via CI parametrisation. Smaller values of
	$\lambda_5$ increases the size of the Yukawa couplings (as evident from Eq.~\eqref{eq:Yuk} and \eqref{eq:zeta}), which in turn would imply an enhancement in the negative contribution to muon $(g-2)$ which is undesirable.
	This can be seen from Fig.~\ref{fig:lam5amu} where the contribution to muon $(g-2)$ (i.e. $\Delta a_\mu$) from the $\eta^+$ and $N_k$ loop as a function of $N_1$ mass is shown keeping other parameters fixed as mentioned in the inset of the figure. The horizontal solid red line depicts a conservative upper limit ($\Delta a_{\mu} = -10^{-10}$) on this negative contribution. Clearly $\lambda_{5}$ smaller than $10^{-10}$ are not allowed from this requirement. We will see similar constraints from LFV also in the next section.    
	
	\subsection{Lepton flavour violation}
	
	Charged lepton flavour violating (CLFV) decay is a promising process to study from beyond standard model (BSM) physics point of view. In the SM, such a process occurs at one-loop level and is suppressed by the smallness of neutrino masses, much beyond the current experimental sensitivity \cite{TheMEG:2016wtm}. Therefore, any future observation of such LFV decays like $\mu \rightarrow e \gamma$ will definitely be a signature of new physics beyond the SM. In the present model, such new physics contribution can come from the charged component of the additional scalar doublet $\eta$ going inside a loop along with singlet fermions, similar to the way it gives negative contribution to lepton $(g-2)$ shown in Fig.~\ref{fig:scotog2}. Adopting the general prescriptions given in \cite{Lavoura:2003xp, Toma:2013zsa}, the decay width of $\mu \rightarrow e \gamma$ can be calculated as
	\begin{align}
		{\rm Br} (\mu \rightarrow e \gamma) =\frac{3 (4\pi)^3 \alpha}{4G^2_F} \lvert A_D \rvert^2 {\rm Br} (\mu \rightarrow e \nu_{\mu} \overline{\nu_e})
	\end{align}
	where $A_D$ is given by
	\begin{equation}
		A_D = \sum_{k} \frac{h^*_{ke} h_{k\mu}}{16 \pi^2} \frac{1}{M^2_{\eta^+}} f (t_k)
		\label{ADMEG}
	\end{equation}
	
	where $t_k = m^2_{N_k}/M^2_{\eta^+}$ and $f(x)$ is the same loop function given by Eq. \eqref{loop1}. The latest bound from the MEG collaboration is $\text{Br}(\mu \rightarrow e \gamma) < 4.2 \times 10^{-13}$ at $90\%$ confidence level \cite{TheMEG:2016wtm}. 
	
	
	\begin{figure}[h!]
		\centering
		\includegraphics[scale=0.5]{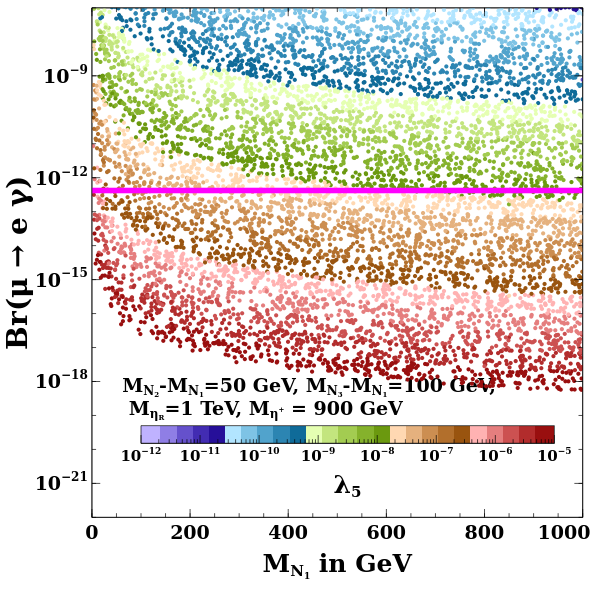}
		\caption{Br$(\mu \rightarrow e \gamma)$ is shown as a function of $N_1$ mass. The colour bar shows the values of $\lambda_{5}$.}
		\label{fig:lam5meg}
	\end{figure}

	In Fig.~\ref{fig:lam5meg}, Br($\mu \rightarrow e \gamma$) is shown as a function of $N_1$ keeping other parameters fixed as mentioned in the inset of the figure. The solid magenta line depicts the latest upper limit from the MEG experiment. It is clear that $\lambda_{5}$ smaller than $\sim \mathcal{O}(10^{-8})$ is disfavoured from the CLFV constraint. 
	\begin{figure}[h!]
		\centering
		\includegraphics[scale=0.5]{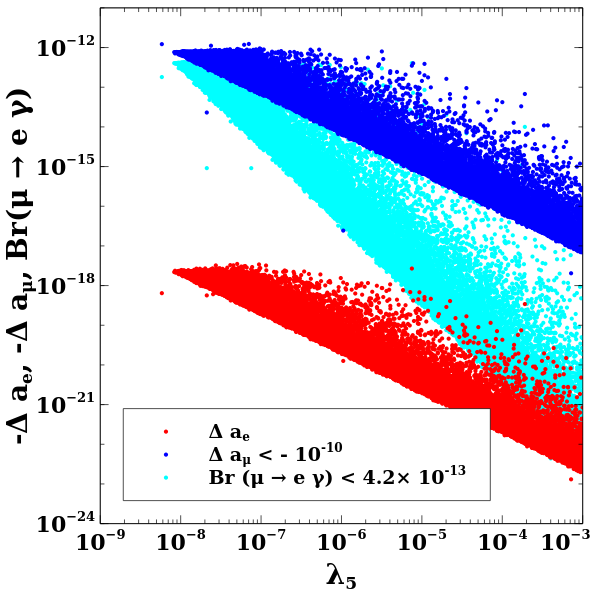}
		\caption{Contribution to (Br$(\mu \rightarrow e \gamma)< 4.2\times10^{-13}$) and ($\Delta a_{\mu}<-10^{-10}$) with $M_1,M_2,M_3 \in [1,1000]$ GeV, $M_{\eta^+} \in [100,1000]$ GeV and $\lambda_{5} \in [10^{-10},10^{-3}]$ in the minimal model. Corresponding values of $\Delta a_e$ are shown by the red coloured points.}.
		\label{fig:lam5cps1}
	\end{figure}
	
	To look for common parameter space satisfying the CLFV constraint from MEG (Br$(\mu \rightarrow e \gamma)< 4.2\times10^{-13}$) and ($\Delta a_{\mu}<-10^{-10}$) from the contribution of $\eta^+$ and $N_k$ in the loop, we carried out a numerical scan varying $M_1,M_2,M_3 \in [1,1000]$ GeV (with $M_1<M_2,M_3$), $M_{\eta^+} \in [100,1000]$ GeV and $\lambda_{5} \in [10^{-10},10^{-3}]$. Along with calculating Br$(\mu \rightarrow e \gamma)$ and $\Delta a_{\mu}$ we also calculate electron $(g-2)$ namely $\Delta a_e$ to check if all of them can be satisfied simultaneously. Only those parameter set which satisfies both the above mentioned constraints (CLFV and $\Delta a_{\mu}<-10^{-10}$) are screened out and the corresponding value of $\Delta a_e$ is noted down for these parameters. The result of this scan is shown in Fig.~\ref{fig:lam5cps1}. Clearly with $\lambda_{5}$ smaller than $10^{-8}$ we can not satisfy the MEG constraint and the constraint on maximum negative contribution to $\Delta a_\mu$ simultaneously. However, for the parameter space satisfying these two constraints, the corresponding values of $\Delta a_e$ are several orders of magnitude below the correct ballpark (i.e $\Delta a_e = (-87 \pm 36) \times 10^{-14}$). 
	
	Clearly, as we increase the value of $\lambda_{5}$, the Yukawa couplings decrease which diminishes $Br(\mu \rightarrow e \gamma)$ as well as $\Delta a_\mu$ coming from the loop diagram given in Fig.~\ref{fig:scotog2}. It is worth mentioning here that, though much smaller values of $\lambda_{5}$ can bring the $\Delta a_e$ to correct ball-park, but such values of $\lambda_5$ will violate the CLFV constraints. Thus it is not possible to explain $\Delta a_{\mu}$ and $\Delta a_e$ simultaneously in this minimal model being within the constraints from LFV. This result is not surprising because the charged scalar loop contribution to electron $(g-2)$ is suppressed by electron mass squared. Presence of additional heavy fermion doublet in the loop can lead to chiral enhancement of $(g-2)_e$, as discussed in earlier works in the context of muon $(g-2)$ \cite{Calibbi:2018rzv, Arcadi:2021cwg}. This is the topic of our next section.

	\section{Extension of minimal scotogenic $U(1)_{L_{\mu}-L_{\tau} }$ model}
	\label{sec3}
	As the minimal model described above can not explain the electron and muon $(g-2)$ data simultaneously while being in agreement with CLFV constraints, we extend the model by introducing a vector like fermion doublet $\Psi^T=(\psi^0, \psi^-) \sim (1, 2, -\frac{1}{2}, 0)$ which is also odd under the $Z_2$ symmetry, but has vanishing $U(1)_{L_{\mu}-L_{\tau} }$ charge. Since there exists a RHN which also has vanishing $U(1)_{L_{\mu}-L_{\tau} }$ charge, this leads to the possibility of singlet-doublet fermion dark matter~\cite{Mahbubani:2005pt,DEramo:2007anh,Enberg:2007rp,Cohen:2011ec,Cheung:2013dua, Restrepo:2015ura,Calibbi:2015nha,Cynolter:2015sua, Bhattacharya:2015qpa,Bhattacharya:2017sml, Bhattacharya:2018fus,Bhattacharya:2018cgx,DuttaBanik:2018emv,Barman:2019tuo,Bhattacharya:2016rqj, Calibbi:2018fqf, Barman:2019aku, Dutta:2020xwn}. Here it is worth mentioning that introduction of a singlet vector like fermion instead of the doublet to the minimal scotogenic $L_{\mu}-L_{\tau}$ model can not lead to the required chiral enhancement in one-loop anomalous magnetic moment of electron \cite{Calibbi:2018rzv, Arcadi:2021cwg}. To make it clear, we show the chirally enhanced one-loop Feynman diagram for electron $(g-2)$ in Fig.~\ref{fig:eg2}. Since we have only one charged scalar $\eta^+$, the requirement of left and right chiral electrons in external legs for chiral enhancement can be fulfilled only with a Higgs VEV insertion in internal fermion line, as clearly seen from the Feynman diagram in the top panel of Fig.~\ref{fig:eg2}. This is possible only when a vector like lepton doublet is introduced to the minimal model discussed before.
	
	With the incorporation of the vector-like fermion doublet $\Psi$, the new terms in the relevant Lagrangian can be written as follows.
	\begin{eqnarray}
		\label{model_Lagrangian}
		\mathcal{L} & = &  \overline{\Psi} \left( i\gamma^\mu D_\mu - M \right) \Psi \nonumber\\&-&{Y_\psi}\overline{\Psi}\widetilde{H}\big(N_{e}+(N_{e})^c\big)-Y_{\psi e} \overline{\Psi_{L}} \eta e_R +{\rm h.c.} 
		\label{sd_lag}    
	\end{eqnarray}
	where $\widetilde{H}=i\tau_2 H^{*}$.   
	${N}_{e}$ being neutral under $U(1)_{L_{\mu}-L_{\tau}}$ has Yukawa coupling with fermion doublet $\Psi$, determining the physical dark matter state of the model after electroweak symmetry breaking (EWSB). Note that we have assumed same Yukawa coupling $Y_{\psi}$ to couple $\Psi$ with $N_e, N^c_e$ for simplicity.
	
	Thanks to the Yukawa interaction in \eqref{sd_lag}, the electromagnetic charge neutral component of 
	$\Psi$ {\it viz.} $\psi^0$ and $N_{e}$ mixes after the SM Higgs acquires a non-zero VEV. The mass terms for these fields can then be written together as follows.
	
	\begin{eqnarray}
		-\mathcal{L}_{\rm mass}& = M\overline{\psi^0_L}\psi^0_R + \frac{1}{2}M_{ee}\overline{N}_{e}(N_{e})^c \nonumber\\&+ {m'_D} (\overline{\psi^0_L}N_{e}+\overline{\psi^0_R}(N_{e})^c) + {\rm h.c.} 
		\label{l_mass}
	\end{eqnarray}
	
	where $m'_D=\frac{{Y_\psi v }}{\sqrt{2}}$, with $ v = 246$ GeV. 
	
	Since $N_e$ mixes with $N_\mu$ and $N_\tau$ (as seen from Eq.~\eqref{eq:yuklag}) apart from its mixing with $\psi^0$, it gives rise to a $5\times5$ mass matrix for the neutral fermions can be written in the basis $((\psi^0_R)^c, \psi^0_L,(N_{e})^c,(N_{\mu})^c,(N_{\tau})^c)^T$ as given in Appendix~\ref{appendix1}.

	As $\psi_L$ and $\psi_R$ has no coupling with $N_\mu$ and $N_\tau$, and $N_e$ can be assumed to be dominantly $N_1$ (one of the the physical RHN states in the absence of singlet-doublet coupling), we ignore the mixing of $\psi^0$ with $N_2$ and $N_3$. This allows us to write down this neutral fermion mass matrix for the dark sector in the basis $ ((\psi^0_R)^c, \psi^0_L, (N_{1})^c)^T$ as :
	\begin{equation}\label{dark-sector-mass}
		\mathcal{M}=
		\left(
		\begin{array}{ccc}
			0 &M &{m_D}\\
			M &0 &{m_D}\\
			{m_D} &{m_D} &M_{1}\\
		\end{array}
		\right)\,.
	\end{equation}
	Note that there is a difference in $m_D$ in \eqref{dark-sector-mass} from $m'_D$ in \eqref{l_mass}, the details of which can be found in Appendix~\ref{appendix1}.
	This symmetric mass matrix $\mathcal{M}$ can be diagonalised by a unitary matrix 
	$\mathcal{U (\theta)}=U_{13}(\theta_{13}=\theta).U_{23}(\theta_{23}=0).U_{12}(\theta_{12}=\frac{\pi}{4})$, which is essentially characterized by a single angle $\theta_{13}=\theta$ \cite{Ghosh:2021khk}. We diagonalise the mass matrix as $\mathcal{M}$ by $\mathcal{U}.\mathcal{M}.\mathcal{U}^T = \mathcal{M}_{\rm Diag}$, 
	where the unitary matrix $\mathcal{U}$ is given by:
	\begin{equation}
		\label{diagonalizing_matrix}
		\mathcal{U}= \left(
		\begin{array}{ccc}
			1 & 0 & 0\\
			0 & e^{i\pi/2} & 0\\
			0 & 0 & 1\\
		\end{array}
		\right)
		\left(
		\begin{array}{ccc}
			\frac{1}{\sqrt{2}}\cos\theta &\frac{1}{\sqrt{2}}\cos\theta &\sin\theta\\
			-\frac{1}{\sqrt{2}} &\frac{1}{\sqrt{2}} &0\\
			-\frac{1}{\sqrt{2}}\sin\theta &-\frac{1}{\sqrt{2}}\sin\theta &\cos\theta\\
		\end{array}
		\right)\\.
	\end{equation}
	The extra phase matrix is multiplied to make sure all the eigenvalues after diagonalisation are positive.

	The diagonalisation of 
	the mass matrix given in Eq.~\eqref{dark-sector-mass} requires
	\begin{equation}
		\tan2\theta = \frac{2\sqrt{2}~ m_D}{M-M_{1}} .
	\end{equation}
	The emerging physical states defined as $\chi_{_i}=\frac{\chi_{_{iL}}+(\chi_{_{iL}})^c}{\sqrt{2}}~(i=1,2,3)$ are 
	related to the flavour or unphysical states as follows.
	\begin{equation}
		\begin{aligned}
			\chi_{_{1L}} & = \frac{\cos\theta}{\sqrt{2}}( \psi^0_L+(\psi^0_R)^c  )+\sin\theta (N_{1})^c,
			\\
			\chi_{_{2L}} & =  \frac{i}{\sqrt{2}}(\psi^0_L - (\psi^0_R)^c), 
			\\
			\chi_{_{3L}} & =  -\frac{\sin\theta}{\sqrt{2}}(\psi^0_L + (\psi^0_R)^c ) +\cos\theta(N_{1})^c \,.
		\end{aligned}
		\label{eq:masseigenstate}
	\end{equation}

	All the three physical states $\chi_{_1}, \chi_{_2} ~{\rm and} ~\chi_{_3}$ are therefore of Majorana nature and their 
	mass eigenvalues are,
	\begin{equation}
		\begin{aligned}
			m_{\chi_{_1}} & = M \cos^2\theta + M_{1} \sin^2\theta +  m_D\sin2\theta,
			\\
			m_{\chi_{_2}} & = M,
			\\
			m_{\chi_{_3}} & = M_{1} \cos^2\theta + M\sin^2\theta - m_D\sin2\theta\,.
			\\
		\end{aligned}
	\end{equation}

	In the small mixing limit ($\theta\to 0$), the eigenvalues can be further simplified as,
	\begin{equation}
		\begin{aligned}
			m_{\chi_{_1}} & \approx M + \frac{ 2\sqrt{2}~m^2_D}{M - M_{1}},
			\\
			m_{\chi_{_2}} & = M,
			\\
			m_{\chi_{_3}} & \approx M_{1} - \frac{2\sqrt{2~}m^2_D}{M - M_{1}}.
			\\
		\end{aligned}
	\end{equation}
	where we have assumed $m_D << M, M_{1}$. Hence it is clear that $m_{\chi_{_1}}>m_{\chi_{_2}}>m_{\chi_{_3}}$ 
	and $\chi_{_3}$, being the lightest, becomes the stable DM candidate. It should be noted that we have considered other $Z_2$ particles not part of the mass matrix \eqref{dark-sector-mass} above to be much heavier. For a detailed analysis of singlet-doublet Majorana DM, one may refer to the recent work~\cite{Dutta:2020xwn}. Since DM is of Majorana nature, diagonal $Z$-mediated interactions are absent marking a crucial difference with the singlet-doublet Dirac DM~\cite{Barman:2019tuo,Bhattacharya:2015qpa,Bhattacharya:2018cgx,Bhattacharya:2016rqj,Bhattacharya:2017sml,Bhattacharya:2018fus}. The relevant Lagrangian along with all possible modes of annihilation and co-annihilations of DM are given in Appendix~\ref{appendix2}.

	By the inverse transformation $\mathcal{U}.\mathcal{M}.\mathcal{U}^T = \mathcal{M}_{\rm Diag}$, we can express $Y_\psi$, $M$ and $M_{1}$ in terms of the physical masses and the mixing angle as,
	\begin{equation}
		\begin{aligned}
			Y_\psi & \approx \frac{\Delta M ~\sin2\theta}{2 v},
			\\
			M & \approx m_{\chi_{_1}} \cos^2\theta +m_{\chi_{_3}} \sin^2\theta, 
			\\
			M_{1} & \approx m_{\chi_{_3}} \cos^2\theta +  m_{\chi_{_1}}\sin^2\theta; 
			\\
		\end{aligned}
		\label{dark_parameters}
	\end{equation}
	where $\Delta M=(m_{\chi_{_1}}-m_{\chi_{_3}})$. We can also see that in the limit of $m_D<<M$, $m_{\chi_{_1}}\approx m_{\chi_{_2}}=M$. The phenomenology of dark sector is therefore governed mainly by the three independent parameters, DM mass ($m_{\chi_3}$), splitting with the Next to the lightest particle ($\Delta M$), and the singlet-doublet mixing parameter $\sin \theta$.

	\subsection{Lepton $(g-2)$ in Extended Model}
	
	The coupling of $\Psi$ with $N_e$ as well as $e_R$ allows the possibility of an chiral enhanced contribution to electron $(g-2)$ unlike in the minimal model discussed earlier. The contribution to muon $(g-2)$ however, remains same as the minimal model at leading order.
	
	The singlet-doublet flavour states can be written in terms of the mass eigenstates by inverting Eq.~\eqref{eq:masseigenstate} as follows.
	\begin{eqnarray}
		(\psi^0_R)^c &=& \frac{\cos\theta}{\sqrt{2}}\chi_{_{1L}}-\frac{1}{\sqrt{2}}\chi_{_{2L}} - \frac{\sin\theta}{\sqrt{2}} \chi_{_{3L}}\nonumber \\
		\psi^0_L &=& \frac{\cos\theta}{\sqrt{2}}\chi_{_{1L}} + \frac{1}{\sqrt{2}}\chi_{_{2L}} - \frac{\sin\theta}{\sqrt{2}} \chi_{_{3L}} \nonumber\\
		(N_1)^c & = & \sin\theta ~\chi_{_{1L}} + \cos\theta ~\chi_{_{3L}} 
	\end{eqnarray}
	
	\begin{figure}[h!]
		\centering
		\includegraphics[scale=0.4]{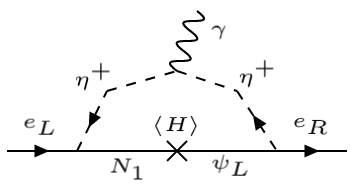}
		\includegraphics[scale=0.3]{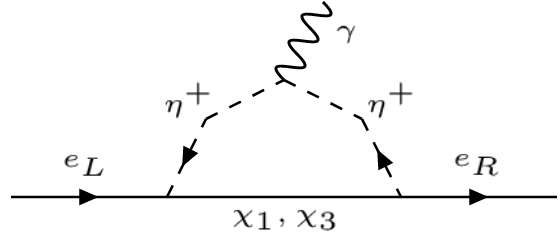}
		\caption{Dominant contribution to electron $(g-2)$ in the extended model. Dark states are shown in flavour basis [top panel] and mass basis [bottom panel] respectively.}
		\label{fig:eg2}
	\end{figure}

	Thus, after writing the flavour states in terms of physical or mass states, one can calculate the $(g-2)$ contribution by considering these physical states $\chi_{_{1,2,3}}$ in the loop. Thus, it is possible to have a chiral enhancement to electron $(g-2)$ as shown in Fig.~\ref{fig:eg2}. Since $N_1$ has no admixture of $\chi_{2}$, so $\chi_{2}$ does not play any role in this loop calculation as it has no coupling with electron. Thus the contribution to $\Delta a_e$ in singlet-doublet model is given by \cite{Calibbi:2018rzv, Jana:2020joi}
	\begin{eqnarray}
		\Delta a_e &=& -\frac{m_e }{8\pi^2 M^2_{\eta^+}}\frac{\sin\theta \cos\theta }{\sqrt{2}}{\rm Re}(h_{1e} Y^*_{\psi e} )\nonumber\\&\times&\Big[m_{\chi_{_1}}f_{LR}\big(\frac{m^2_{\chi_{1}}}{M^2_{\eta^+}}\big)-m_{\chi_{_3}}f_{LR}\big(\frac{m^2_{\chi_{3}}}{M^2_{\eta^+}}\big)\Big] 
	\end{eqnarray}
	where
	\begin{equation}
		f_{LR}(x)=\frac{1-x^2+2x\log{x}}{2 (1-x)^3}
		\label{loop2}
	\end{equation}
	\begin{figure}[h!]
		\centering
		\includegraphics[scale=0.5]{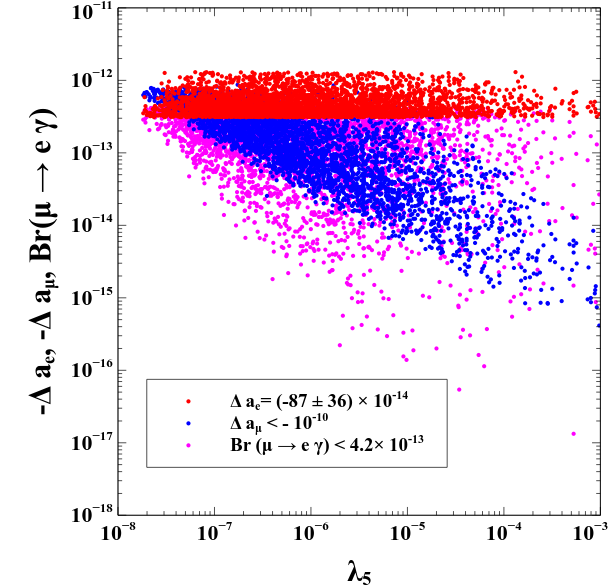}
		\caption{Contribution to $\Delta a_e$ a well as (Br$(\mu \rightarrow e \gamma)< 4.2\times10^{-13}$) and ($\Delta a_{\mu}<-10^{-10}$) in the extended model.}
		\label{fig:lam5cps2}
	\end{figure}
	In Fig.~\ref{fig:lam5cps2}, the result of a parameter scan similar to Fig.~\ref{fig:lam5cps1} is shown after incorporating the additional and dominant contribution to $\Delta a_e$ from $\chi_{1}$ and $\chi_{3}$ present in the singlet-doublet model. For this scan $m_{\chi_{3}}$ and $\Delta m$ are randomly varied in the range $m_{\chi_{3}} \in [1,1000]$ GeV and $\Delta m \in [1,100]$ GeV respectively. Similar to the minimal model, the charged scalar mass is varied as $M_{\eta^+} \in [100,1000]$ GeV. The other two parameters which are randomly varied are $\sin\theta \in [0.01,1]$ and $Y_{\psi e} \in [10^{-2},1]$. 
	 
\begin{figure}[h!]
		\centering
		\includegraphics[scale=0.5]{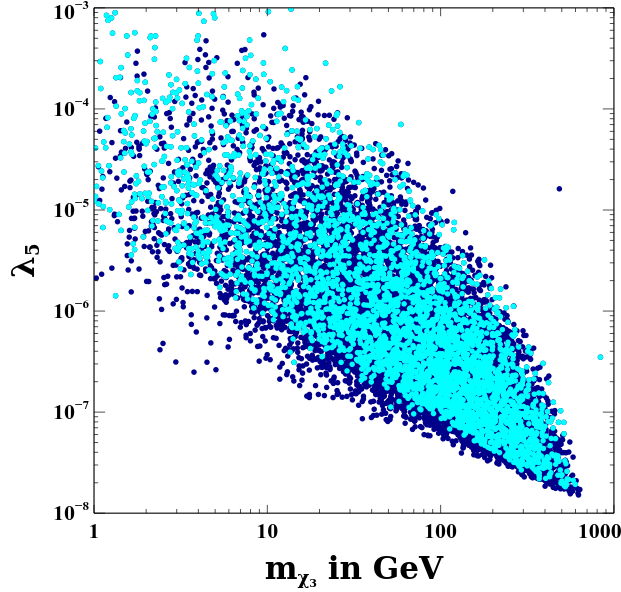}
		\caption{Common parameter space giving correct $\Delta a_e$ as well as satisfying (Br$(\mu \rightarrow e \gamma)< 4.2\times10^{-13}$) and ($\Delta a_{\mu}<-10^{-10}$) in the plane of $\lambda_{5}$ and $m_{\chi_{3}}$.}
		\label{fig:final_param}
	\end{figure}

It is worth mentioning here that, in this extended frame work, along with $(g-2)$ of electron, the LFV process $\mu \to e \gamma$ can also get an chiral enhancement because of the off-diagonal structure of the Yukawa matrix obtained through Casas-Ibarra parametrisation. This chirally enhanced contribution to $\mu \to e \gamma$ amplitude and decay rate are given by
\begin{eqnarray}
			A &=& \frac{1 }{32\pi^2 M^2_{\eta^+}}\frac{\sin\theta \cos\theta }{\sqrt{2}}{\rm Re}(h_{1\mu} Y^*_{\psi e} )\nonumber\\&\times&\Big[2 \frac{m_{\chi_{_1}}}{m_\mu} f_{LR}\big(\frac{m^2_{\chi_{1}}}{M^2_{\eta^+}}\big)-2\frac{m_{\chi_{_3}}}{m_\mu}f_{LR}\big(\frac{m^2_{\chi_{3}}}{M^2_{\eta^+}}\big)\Big], \nonumber
		\end{eqnarray}
		\begin{eqnarray}
			{\rm Br}(\mu \to e \gamma) &=& \tau_{\mu} \frac{\alpha_{em} m^5_{\mu}}{4} |A|^2,
		\end{eqnarray}
	with $\tau_{\mu}$ being the lifetime of muon.
After incorporating this contribution, Only those parameter sets which satisfy all the three constraints i.e. (Br$(\mu \rightarrow e \gamma)< 4.2\times10^{-13}$), ($\Delta a_{\mu}<-10^{-10}$)  and ($\Delta a_e = (-87 \pm 36) \times 10^{-14}$) simultaneously are screened out to get the common parameter space which are shown in Fig.~\ref{fig:lam5cps2}. Clearly, the red points in Fig.~\ref{fig:lam5cps2} depict that in the extended model we can obtain a parameter space that gives $\Delta a_e$ in the correct ballpark as suggested by the experiments while being within the limits of CLFV and suppressed negative contribution to $\Delta a_\mu$. The final parameter space giving correct $\Delta a_e$ as well as satisfying CLFV constraint and ($\Delta a_{\mu}<-10^{-10}$) is shown in Fig.~\ref{fig:final_param} in the plane of $\lambda_{5}$ and $m_{\chi_{3}}$. The blue points are obtained before incorporating the chirally enhanced contribution to $\mu \to e \gamma$ and the cyan points depict the parameter sets which satisfy all the three above-mentioned constraints even after including the chiral enhancement to $\mu \to e \gamma$ in the calculation. Clearly, the enhancement to this CLFV process slightly reduces the allowed parameter space.


	\section{Dark Matter Phenomenology}
	\label{sec4}

	
	As mentioned earlier, we consider the singlet-doublet fermion DM scenario in our work. Before proceeding to calculate the DM relic density numerically, let us first study the possible dependence of DM relic on important relevant parameters namely, the 
	mass of DM ($m_{\chi_{3}}$), the mass splitting ($\Delta M$) between the DM $\chi_{_3}$ and the next-to-lightest stable 
	particles (NLSP) ($m_{\chi_{2}} \approx m_{\psi^{\pm}} \approx m_{\chi_{1}}$) and the mixing angle $\sin \theta$. Depending on the relative magnitudes of some of these parameters, DM relic can be generated dominantly by annihilation or co-annihilation or a combination of both. Effects of co-annihilation on DM relic density, specially when mass splitting between DM and NLSP is small, has been discussed in several earlier works including \cite{Griest:1990kh, Edsjo:1997bg}.

\begin{figure}[h!]
		\centering
		\includegraphics[width = 90 mm]{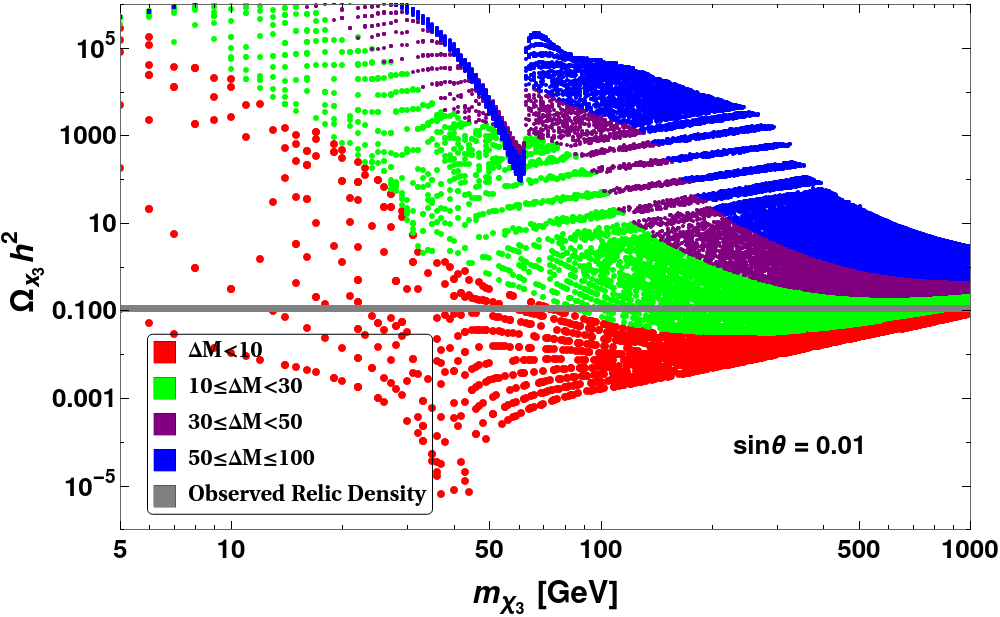}
		\includegraphics[width = 90 mm]{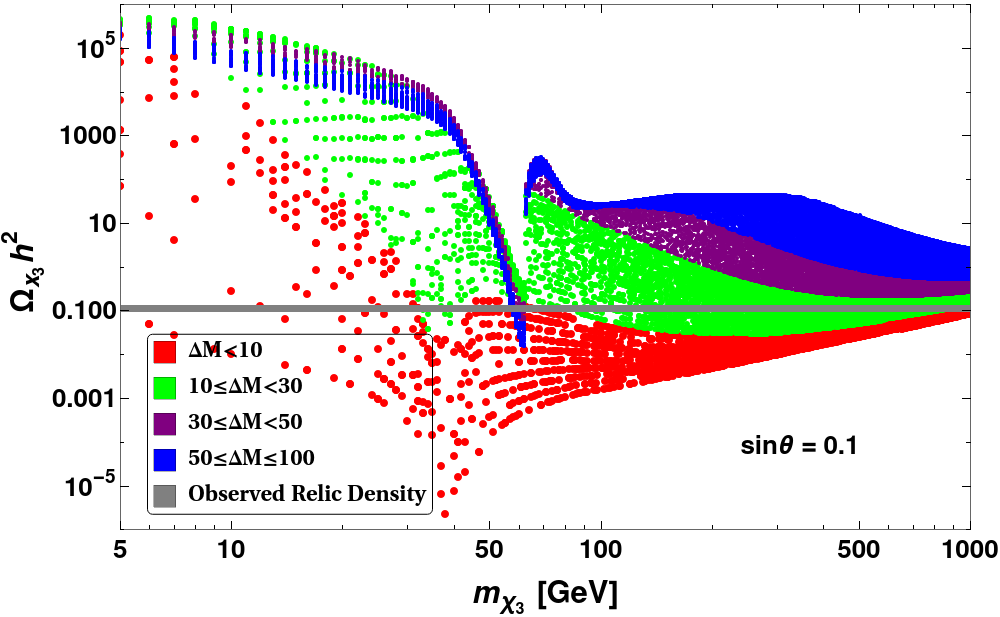}
		\includegraphics[width = 90 mm]{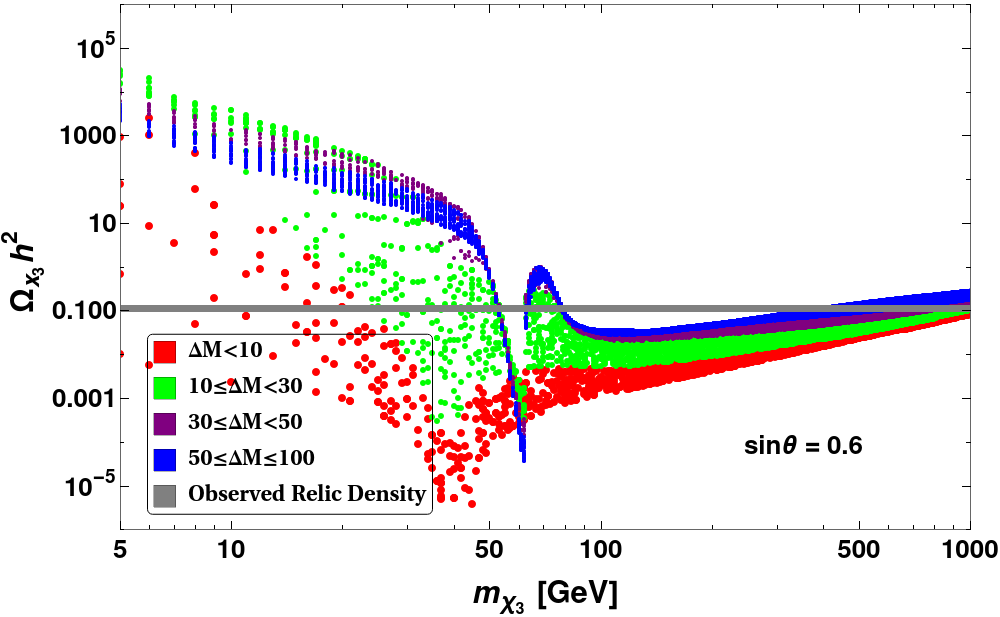}
		
		\caption{DM relic density as a function of DM mass ($m_{\chi_3}$) for different mass splitting $\Delta M$ between the DM and the NLSP (as mentioned in figure inset in GeV) 
			for $\sin\theta=0.01$ (top panel), $\sin\theta=0.1$ (middle panel),  $\sin\theta=0.6$ (bottom panel). 
			Correct relic density region from Planck 2018 data ($\Omega_{\rm DM} h^2 = 0.120 \pm 0.001$) \cite{Aghanim:2018eyx} is indicated by the grey coloured horizontal solid line.}
		\label{fig:relicsplot1}
	\end{figure}

	We adopt a numerical way of computing annihilation cross-section and relic 
	density by implementing the model into the package {\tt MicrOmegas}~\cite{Belanger:2008sj}, where the model files are generated 
	using {\tt FeynRule}~\cite{Christensen:2008py, Alloul:2013bka}. Variation of relic density of DM $\chi_{_3}$ is shown in Fig.~\ref{fig:relicsplot1} 
	as a function of its mass for different choices of $\Delta M$ = 1-10 GeV, 10-30 GeV, 30-50 GeV, 50-100 GeV shown by 
	different colour shades as indicated in the figure inset. The mixing angle is assumed to take values $\sin\theta $= 0.01 (top panel), 0.1 (middle panel) and 0.6 (bottom panel).

	As it can be seen from Fig.~\ref{fig:relicsplot1}, when $\Delta M$ is small, relic density is smaller due to large co-annihilation 
	contribution from $W^\pm$ mediated and flavour changing $Z$-mediated processes. As co-annihilation effects increase, we notice enhanced resonance effect as expected. As these interactions are off-diagonal, the resonances are somewhat flattened compared to a sharp spike expected for diagonal interactions. As $\Delta M$ increases, these co-annihilations become less and less effective, and Higgs mediated annihilations starts dominating. For $\Delta M=30~ {\rm GeV}$, both contributions are present in comparable amount while for $\Delta M> 30~{\rm GeV}$, the contributions from gauge boson mediated (co-annihilation) interactions are practically negligible and the the Higgs mediated channels dominate. Consequently,  a resonance at SM-Higgs threshold $m_{\chi_3} \sim m_{h}/2$ appears, while the the same at $m_{\chi_3} \sim m_{Z}/2$ disappears. It is also observed that as long as $\Delta M$ is small and the co-annihilation channels dominate, the effect of $\sin\theta$ on relic density is negligible. For smaller 
	$\sin \theta$, the annihilation cross-section due to Higgs portal is small leading to larger relic 
	abundance, while for large $\sin \theta$, the effective annihilation cross-section is large leading to smaller relic abundance. 
	However, this can only be observed when $\Delta M$ is sufficiently large enough and the effect of co-annihilation is negligible. 
	In Fig.~\ref{fig:relicsplot1}, the correct DM relic density ($\Omega_{\rm DM} h^2 = 0.120 \pm 0.001$) from Planck 2018 data~\cite{Aghanim:2018eyx} is shown by the grey coloured horizontal solid line. Note that in Fig.~\ref{fig:relicsplot1}, we have chosen $M_{Z_{\mu \tau}}=0.2$ GeV, $g_{\mu \tau }= 5 \times 10^{-4}$, kinetic mixing parameter between $U(1)_Y$ of SM and $U(1)_{L_{\mu}-L_{\tau}}$, $\epsilon=g_{\mu \tau}/70$, mass of $L_\mu - L_\tau$-like Higgs, $m_{h_{2,3}}=900$ GeV and the their mixing with SM Higgs to be very small ($\sin\beta_{2,3}=0.003$), consistent with the available constraints. Due to small coupling the effects of annihilation and co-annihilation processes involving $Z_{\mu \tau}$ and $h_{2,3}$ are negligible. Also note that the co-annihilation effect of the inert doublet $\eta$ is effective only when its mass is very close to the DM mass and the corresponding Yukawa couplings $Y_{De}$ and $Y_{\psi e}$ are sizeable. We keep $m_\eta-m_\chi \geq 100 {\rm GeV}$ while the size of the relevant Yukawa couplings very small ($~10^{-3}$) in our analysis and hence it does not affect DM relic density significantly. 
\begin{figure}[h!]
		\centering
		\includegraphics[width = 90 mm]{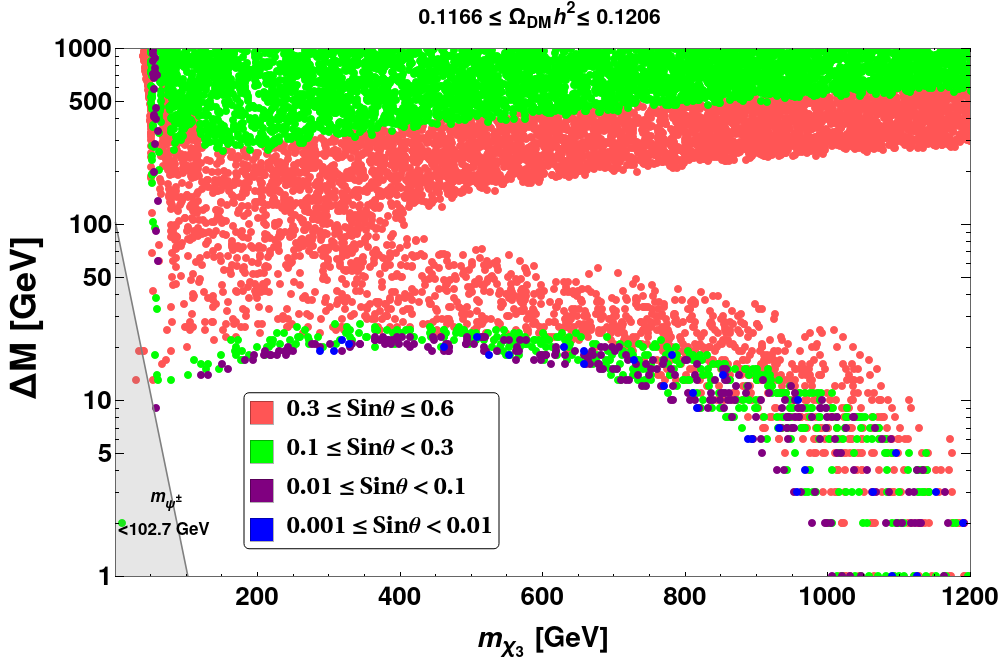}
		\includegraphics[width = 90 mm]{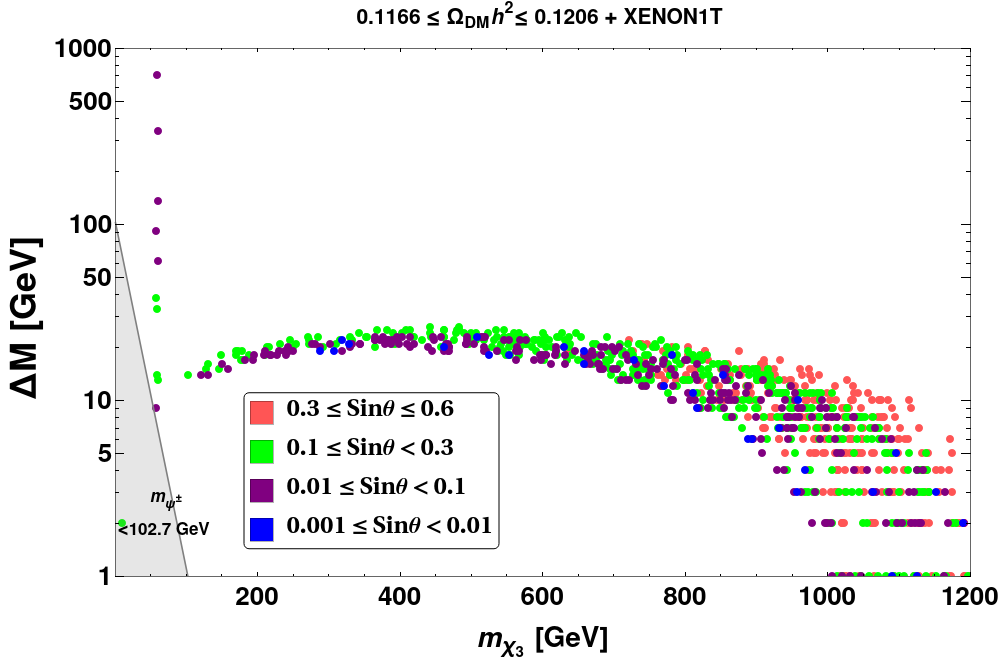}
		
		\caption{[Top panel]: DM relic density (from Planck) allowed parameter space, 
		[Bottom panel]: relic density (from Planck)+ Direct Search (from XENON1T) allowed parameter space in the $\Delta M $ versus $m_{\chi_3}$ plane for different ranges of $\sin\theta$. Shaded region in the bottom left corner is ruled out by LEP exclusion bound on charged fermion mass,$m_{\psi^\pm}=M>102.7$GeV.}
		\label{relic3}
	\end{figure}

	In top panel of Fig.~\ref{relic3}, the 
	correct relic density allowed parameter space has been shown in the plane of $\Delta M$ versus $m_{\chi_{_3}}$ for a wide range of values for the 
	mixing angle $\{\sin\theta=  0.001-0.01, 0.01-0.1, 0.1-0.3,0.3-0.6\}$, indicated by different colours as shown in the figure inset. Note that the chosen ranges of $\Delta M$ as well as mixing angle $\theta$ keep the relevant Yukawa coupling perturbative $Y_\psi < \sqrt{4\pi}$, as seen from Eq. \eqref{dark_parameters}.
	We can see from the top panel of Fig.~\ref{relic3} that there is a bifurcation around $\Delta M \sim 50$ GeV, so that the allowed plane of $m_{\chi_{_3}}-\Delta M$ are separated into two regions: 
	(I) the bottom portion with small $\Delta M$ ($\Delta M \leq$ 50 GeV), where $\Delta M$ decreases with larger DM mass ($m_{\chi_{_3}}$) and (II) the top portion with large $\Delta M$ ($\Delta M \geq$ 50 GeV), where $\Delta M$ increases slowly with larger DM mass $m_{\chi_{_3}}$. In order to understand this figure and two regimes (I) and (II), we note the following.
	
	$\bullet$ In region (I), for a given $\sin\theta$ range, the annihilation cross-section decreases with increase in DM mass 
	$m_{\chi_{_3}}$ and hence more co-annihilation contributions are required to get the correct 
	relic density, resulting $\Delta M$ to decrease. So the region below each of these coloured 
	zones corresponds to under-abundant DM (small $\Delta M$ implying large co-annihilation for a given $m_{\chi_{_3}}$), while the region 
	above corresponds to over-abundant DM due to the the same logic. In this region the Yukawa coupling $Y_\psi$ which governs the annihilation cross-section is comparatively small since $Y_\psi \propto \Delta M \sin\theta$ and $\Delta M$ is small. Also the annihilation cross-section decreases with increase in DM mass. Therefore, when DM mass is sufficiently heavy ($m_{\chi_3} > 1.2$ TeV), annihilation becomes too weak to be compensated by the co-annihilation even when $\Delta M \rightarrow 0$, producing DM over-abundance\footnote{However, $\Delta M$ can not be arbitrarily small as with $\Delta M \to 0$, the charged companions $\psi^{\pm}$ are degenerate with DM and are stable. We can put a lower bound on $\Delta M$ by requiring the charged partners $\psi^{\pm}$ of the 
		DM to decay before the onset of Big Bang Nucleosynthesis ($ \tau_{\rm BBN} \sim 1$ sec.). One may refer to~\cite{Dutta:2020xwn} for further details.}.
	
	$\bullet$ In region (II), the co-annihilation contribution to relic is negligible, thanks to large $\Delta M$. Therefore, Higgs-mediated annihilation processes dominantly contribute to the relic density. As Higgs Yukawa coupling $Y_\psi \propto \Delta M \sin2\theta$, for a given $\sin\theta$, larger $\Delta M$ leads to larger $Y_\psi$ and hence larger annihilation cross-section to yield DM under-abundance, which can only be brought back to the correct ballpark by having a larger DM mass. By the same logic, larger $\sin\theta$ requires smaller $\Delta M$. Therefore, the region above each coloured zone (giving correct relic density for a specific range of $\sin\theta$) is under-abundant, while the region below each coloured zone is over-abundant.
	
	Now imposing the constraints from DM direct search experiments on top of the relic density allowed parameter space (top panel of Fig.~\ref{relic3}) in the $\Delta M$ versus $m_{\chi_3}$ plane, we get the bottom panel of Fig.~\ref{relic3}, which is crucially tamed down as compared to the only relic density allowed parameter space. Here we consider elastic scattering of the DM off nuclei via 
	Higgs-mediated interaction and confront our calculated value of direct search cross-section with that from XENON1T~\cite{Aprile:2018dbl}. Again, the absence of tree level Z-mediated direct search channel makes a crucial difference in the direct search allowed parameter space as compared to singlet-doublet Dirac fermion DM as elaborated in~\cite{Bhattacharya:2017sml,Barman:2019tuo,Bhattacharya:2018cgx,Bhattacharya:2016rqj,Bhattacharya:2015qpa,Bhattacharya:2018fus}. While a large $\sin \theta$ (upto 0.6) is allowed in the present case simultaneously by relic as well as direct search, only upto $\sin \theta \sim 0.01$ is allowed in case of singlet-doublet Dirac fermion DM. The cross section per nucleon for the spin-independent 
	(SI) DM-nucleon interaction is then given by~\cite{Dutta:2020xwn}
	\begin{equation}
		\label{dda2}
		\begin{aligned}
			\sigma_{\rm SI} &= \frac{4}{\pi A^2}\mu^2_r\frac{Y_\psi^2 \sin^2 2\theta}{M^4_h}\Big[\frac{m_p}{v}\Big(f^{p}_{Tu} + f^{p}_{Td} + f^{p}_{Ts} + \frac{2}{9}f^{p}_{TG}\\
			&+\frac{m_n}{v}\Big(f^{n}_{Tu} + f^{n}_{Td} + f^{n}_{Ts} + \frac{2}{9}f^{n}_{TG}\Big)\Big]^2
		\end{aligned}
	\end{equation}
	where A is the mass number of Xenon nucleus, $m_p (m_n)$ is mass of proton (neutron) and $M_h$ is mass of the SM Higgs boson\footnote{Different coupling strengths between DM and light quarks are given by \cite{Bertone:2004pz,Alarcon:2012nr} as $f^p_{Tu} = 0.020 \pm 0.004, f^p_{Td} = 0.026 \pm 0.005, f^p_{Ts} = 0.014 \pm 0.062$, $f^n_{Tu} = 0.020 \pm 0.004, 
		f^n_{Td} = 0.036 \pm 0.005, f^n_{Ts} = 0.118 \pm 0.062$. The coupling of DM with the gluons in target nuclei is parametrised by $
		f^{(p,n)}_{TG} = 1- \sum_{q=u,d,s}f^{p,n}_{Tq}$. See \cite{Hoferichter:2017olk} for more recent estimates.}.

The specific reasons for direct search constraints to rule out heavy fermion mixing is due to the explicit presence of the factor $Y^2_\psi \sin^2 2\theta$ in the direct search cross-section given by Eq. \eqref{dda2}, where $Y_\psi = \Delta M ~\sin2\theta/2 v$ according to Eq. \eqref{dark_parameters}. So the overall dependency of the direct search cross-section on the mass splitting and the singlet-doublet mixing goes as, $\sigma_{\rm SI} \sim \Delta M^2 \sin^42\theta $. Definitely, combination of large $\Delta M$ and large $\sin \theta$ will not survive the direct search bound. Note that relic density favours larger $\sin\theta$ with large $\Delta M$ in order to be within the Planck limit by virtue of large annihilation. So the region roughly above  $\Delta M =$20 GeV can not simultaneously satisfy both the bounds.
The region roughly below $\Delta M$= 20 GeV is perfectly allowed by direct search even for large $\sin \theta$. But from the relic point of view, direct search allowed points with large $\sin \theta$ would lead to under-abundance for DM mass upto $m_{\chi_3}\sim $ 700 GeV due to large co-annihilation rates. However, the region beyond $m_{\chi_3}\sim$ 700 GeV is allowed since annihilation also decreases with increase in DM mass, which compensate for the increase in co-annihilation, giving the correct relic. When we consider both relic density and direct search constraints simultaneously as shown in the bottom panel of Fig. \ref{relic3}, large $\sin \theta$ is allowed only towards higher DM mass with smaller $\Delta M$ favouring a degenerate DM spectrum. The SM Higgs resonance $m_{\chi_3}\sim m_h/2$ is seen to satisfy both relic density and direct search bound, where $\Delta M$ can be very large having very small $\sin\theta$.
		
		
		\begin{figure}[h!]
			\centering
			\includegraphics[scale=0.2]{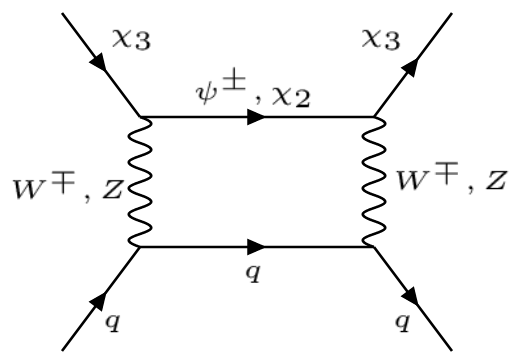}\hfil
			\includegraphics[scale=0.2]{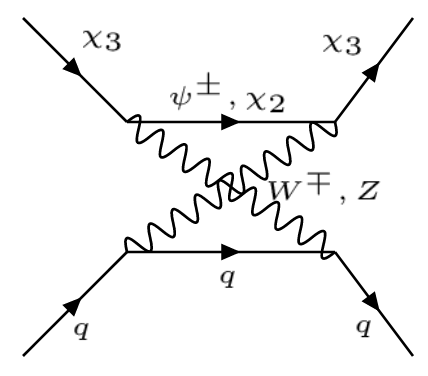}
			\caption{ Spin-independent elastic DM-nucleon scattering arising from the loop exchange of the vector mediators.}
			\label{loopdd}
		\end{figure} 
		
		\begin{figure}[h!]
			\centering
			\includegraphics[scale=0.45]{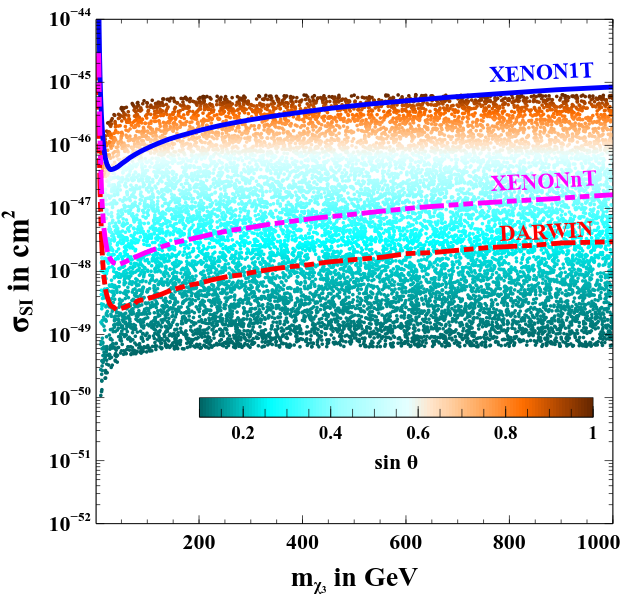}
			\caption{ Loop-induced SI direct detection cross-section as a function of DM mass.}
			\label{lidd}
		\end{figure} 
		
	In addition to the tree level t-channel process for the direct detection prospect of DM discussed here, another contribution to spin-independent direct search cross section can be induced via the electro-weak couplings at loop level~\cite{Bell:2018zra}. The corresponding Feynman diagram is shown in Fig.~\ref{loopdd}. 

%
		   	
This cross-section is given by:
				\begin{equation}
						\label{ddloop}
						\begin{aligned}
								\sigma_{\rm SI} &= \frac{1}{\pi A^2}\mu^2_r \vert \mathcal{M} \vert^2
							\end{aligned}
					\end{equation}	
			where the amplitude is given by
		\begin{equation}
			\mathcal{M}= \frac{4 g^4 m_N m_{\chi_{3}}}{16 \pi^2 M^4_V} F\left(\frac{m^2_{\chi_{3}}}{M^2_V}\right) \sin^2\theta \left[Z f_p + (A-Z) f_n\right]
		\end{equation}
		and the loop function $F$ is given by:
		\begin{equation}
				\label{amp}
				\begin{aligned}
						F(x)=&\frac{(8 x^2-4x+2)\log[\frac{\sqrt{1-4x}+1}{2\sqrt{x}}]}{4 x^2 \sqrt{1-4x}}\\+&\frac{\sqrt{1-4x}(2x+\log(x))}{4 x^2 \sqrt{1-4x}}
					\end{aligned}
			\end{equation}		
			In the above expression $\mu_r$ is the reduced mass and $M_{V}$ is the mass of SM vector boson ($W^\pm$ or $Z$) and $f_p$ and $f_n$ are the interaction strengths (including hadronic uncertainties) of DM with proton and neutron
			respectively. For simplicity we assume conservation of isospin, {\it i.e.} $f_p/f_n = 1$. The value of $f_n$ vary within a range of $0.14<f_n<0.66$ and we take the central value $f_n \simeq 1/3$ ~\cite{Mei:2018qnt,Bhattacharya:2017sml}. In Fig.~\ref{lidd}, we have shown this loop induced spin-independent DM-nucleon scattering cross-section as a function of DM mass $m_{\chi_{3}}$ where the color code represents the value of singlet-doublet mixing $\sin\theta$. It is evident from this figure that, this loop induced DM-nucleon scattering cross-section is almost independent of DM mass consistent with the result presented in ~\cite{Cirelli:2005uq}. 
Clearly, large values of singlet-doublet mixing, $\sin\theta > 0.8$ are ruled out by the latest constraint from the XENON1T experiment (shown by the blue solid line) for DM mass below $600$ GeV. However, further smaller values of mixing angle can be probed by the future experiments like XENONnT and DARWIN, the sensitivity of which are shown by the magenta and red dotted lines respectively. 	
		   	
	Note that although there is a possibility of direct detection by electron recoil though $Z-Z_{\mu \tau}$ mixing by assuming a sub GeV or GeV scale DM either through elastic~\cite{Borah:2021yek} or inelastic scattering~\cite{Okada:2019sbb, Dutta:2021wbn}, the corresponding relic density for such a sub-GeV DM will be over-abundant by several orders of magnitude.

	\begin{figure}[h!]
		\centering
		\includegraphics[scale=0.3]{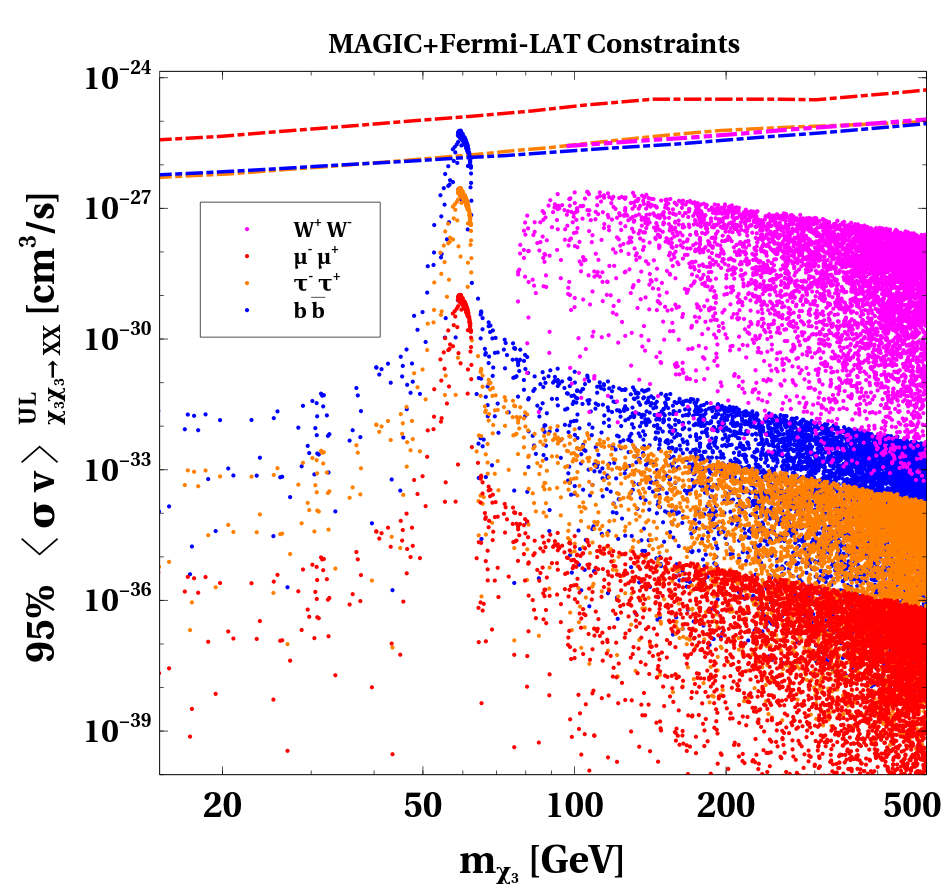}
		\caption{$\langle\sigma v\rangle_{\chi_{3}\chi_{3}\to XX}$ are shown as a function of DM mass where $X$ is the species as mentioned in the inset of figure. Only the points that satisfy DM relic and direct detection constraint are shown.}
		\label{ind_det}
	\end{figure}
DM in WIMP paradigm can also be probed by different indirect detection experiments which essentially search for SM particles produced through DM annihilations. Among these final states, photon and neutrinos, being neutral and stable can reach the indirect detection experiments without getting affected much by intermediate medium. These photons, which are produced from electromagnetically charged final states, lie in the gamma ray regime for typical WIMP DM and hence can be measured at space-based telescopes like the Fermi Large Area Telescope (Fermi-LAT) or ground based telescopes like MAGIC or HESS. Measuring the gamma ray flux and using the standard astrophysical inputs, one can constrain the DM annihilation into different final states like $W^{+}W^{-}$,$\mu^{+} \mu^{-}$, $\tau^{+} \tau^{-}$,$b\bar{b}$. In Fig.\ref{ind_det}, we show the points satisfying both relic constraint and direct search constraint confronted with the combined constraints from MAGIC and Fermi-LAT\cite{MAGIC:2016xys} for annihilation of DM into different species as mentioned in the inset of figure. The dotted lines of different colours show the corresponding upper limit on the DM annihilation cross-section from MAGIC+Fermi-LAT. The more recent analysis indicate similar upper bounds \cite{Alvarez:2020cmw, HESS:2018cbt}. Clearly, a small part of the parameter space, near SM Higgs resonance region, can be disfavoured while future measurements can be sensitive to some parts of the heavier DM mass regime.

	\section{Collider Signatures}
	\label{collider}
	Thanks to the presence of the doublet $\Psi$, the singlet-doublet model has attractive collider signatures such as- Opposite sign dilepton + missing energy $(\ell^+ \ell^- +\slashed{E_T})$, three leptons + missing energy $(\ell \ell \ell +\slashed{E_T})$ etc, see~\cite{Dutta:2020xwn, Bhattacharya:2021ltd}. While such conventional collider signatures have been discussed in details in earlier works, here we briefly comment on an interesting feature of the model: the possibility of displaced vertex signature of $\psi^{\pm}$. Once these particles are produced at colliders by virtue of their electroweak gauge interactions, they can live for longer period before decaying into final state particles including DM \cite{Bhattacharya:2017sml,Bhattacharya:2018fus, Borah:2018smz}. A particle like $\psi^{\pm}$ (which is the NLSP in our model) with sufficiently long lifetime, so that its decay length is of the order of 1 mm or longer, if produced at the colliders, can leave a displaced vertex signature. Such a vertex which is created by the decay of the long-lived particle, is located away from the collision point where it was created. The final state like charged leptons or jets from such displaced vertex can then be reconstructed by dedicated analysis, some of which in the context of the Large hadron collider (LHC) may be found in \cite{ATLAS:2016tbt, CMS:2016kce, Aaboud:2017iio}. Similar analysis in the context of upcoming experiments like MATHUSLA, electron-proton colliders may be found in \cite{Curtin:2017izq, Curtin:2017bxr} and references therein.

\begin{figure}[h!]
		\centering
		\includegraphics[height=8.0cm]{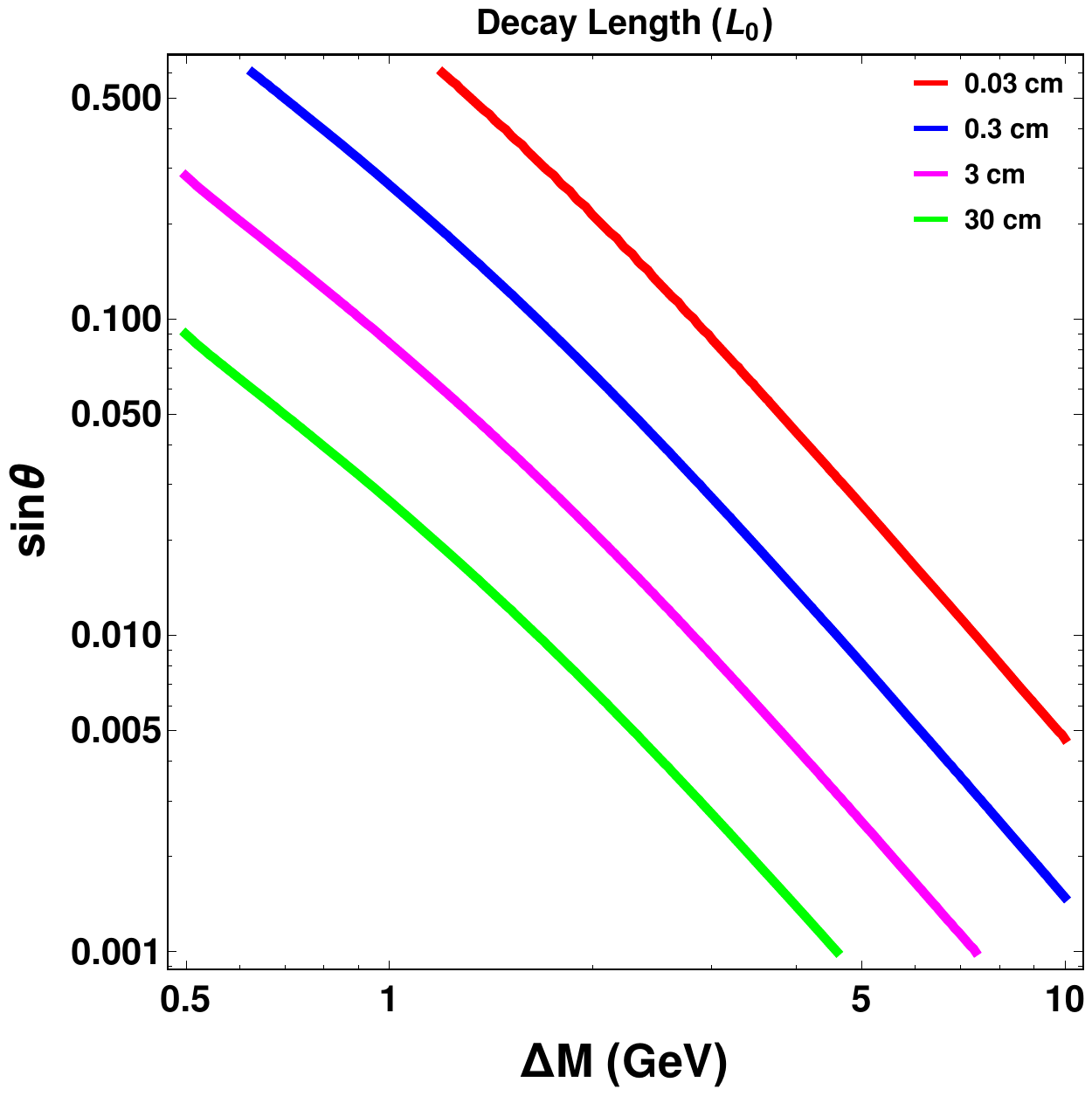}
		\caption{Contours of decay length ($L_0$) of $\psi^\pm$ in the $\Delta M-\sin \theta$ plane.}
		\label{DV_log}
	\end{figure}

	Since a large region of available parameter space of the model relies on small $\Delta M$ (see in the bottom panel of Fig.~\ref{relic3}), the decay of $\psi^\pm$ may be phase space suppressed and can produce very interesting displaced vertex signature. The decay rate for the allowed processes $\psi^\pm \rightarrow \chi_{_3} \pi^\pm$ and $\psi^\pm \rightarrow \chi_{_3} l^{\pm} \nu_{_l}$ in the limit of small $\Delta M$ is given by
	\begin{equation}
		\begin{aligned}
			\Gamma_{\psi^\pm \to \chi_{_3} \pi^\pm} \approx \frac{G^2_F}{\pi}(f_\pi \cos\theta_c)^2 \sin^2\theta ~\Delta M^3 \sqrt{1-\frac{m^2_{\pi^\pm}}{\Delta M^2}},\\
			\Gamma_{\psi^\pm \to \chi_{_3} l^{\pm} \nu_{_l}} \approx \frac{G^2_F}{15 \pi^3}\sin^2\theta~ \Delta M^5 \sqrt{1-\frac{m^2_l}{\Delta M^2}},
		\end{aligned}
		\label{eq:decay_lv}
	\end{equation} 
	where $G_F = 1.16 \times 10^{-5} \; {\rm GeV}^{-2}$ is the Fermi constant, $f_\pi \approx 135$ MeV is the pion form factor, $\theta_c$ is the Cabibbo angle and $\sin \theta$ is the singlet-doublet mixing angle. Using the decay width given by Eq.~\eqref{eq:decay_lv}, we can calculate the decay length $L_0$ of $\psi^\pm$ in the rest frame of $\psi^\pm$. In Fig.~\ref{DV_log}, we show the contours of decay length $L_0$ in the  $\Delta M-\sin\theta$ plane, considering $\Delta M$ and $\sin \theta$ in the range allowed by all other constraints related to DM, $(g-2)$ as well as CLFV. We see that, for sufficiently small $\Delta M  (\Delta M < 10$ GeV), the decay length ($L_0$) can be significantly large to be detected at the collider, while non-observation of a displaced vertex or a charge track will result in a bound on $\Delta M-\sin\theta$ plane.  If the mass splitting is even smaller, say of the order of $\mathcal{O}(100\; \rm MeV)$, the decaying particle $\psi^{\pm}$ can be long-lived enough to give rise to disappearing charged track signatures \cite{Borah:2018smz, Biswas:2018ybc, Biswas:2019ygr} which are also constrained by the LHC \cite{ATLAS:2017oal}. We do not discuss this possibility here and refer to the above-mentioned works and references therein for further details.
	
	\section{Summary and Conclusion}
	\label{sec6}
\begin{figure}[h!]
		\centering
		\includegraphics[scale=0.5]{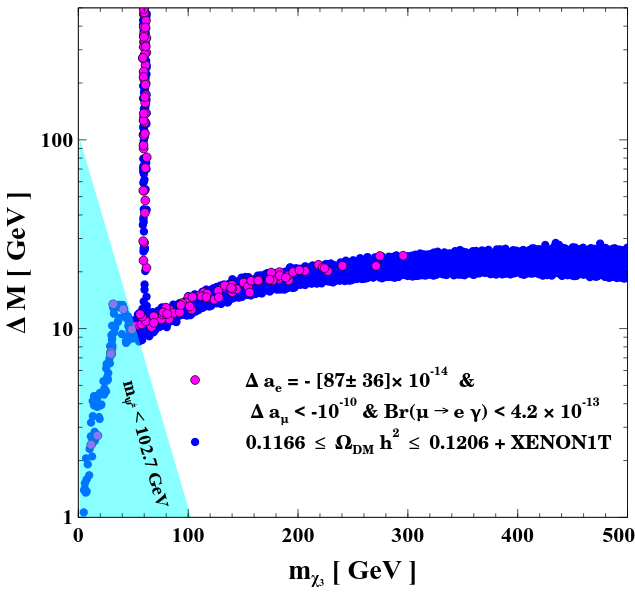}
		\caption{Final parameter space simultaneously allowed from $\Delta a_{e}=- [87\pm 36]\times 10^{-14}$, $\Delta a_{\mu}<-10^{-10}$ \& Br($\mu \rightarrow e \gamma$)$<4.2\times10^{-13}$ as well as constraints from correct relic density DM and direct search of DM at XENON1T. Cyan shaded region is ruled out by LEP exclusion bound on charged fermion mass.}
		\label{finalparam}
	\end{figure}

	Motivated by the growing evidences for anomalous magnetic moment of muon together with recent hints of electron anomalous magnetic moment, but in the opposite direction compared to muon, we study a well motivated particle physics scenario based on gauged $L_{\mu}-L_{\tau}$ symmetry. While the minimal model does not have any dark matter candidate but explains light neutrino masses via type I seesaw mechanism at tree level, there exists a small parameter space currently allowed from all limits which is consistent with observed muon $(g-2)$ where the positive contribution to $(g-2)$ comes from light vector boson loop. In order to accommodate DM and a negative electron $(g-2)$, we first consider a scotogenic extension of the model by including an additional scalar doublet $\eta$ and an in-built $Z_2$ symmetry under which RHNs and $\eta$ are odd while SM fields are even. Even though there exists a charged scalar loop contribution to $(g-2)$ in this model, due to the absence of chiral enhancement, it is not possible to explain $(g-2)_e$ while being consistent with overall positive $(g-2)_{\mu}$ and other bounds from neutrino mass, LFV etc. Therefore, we further extended the model by an additional vector like lepton doublet to get an enhanced negative contribution to electron $(g-2)$ with DM phenomenology driven by the well-studied singlet-doublet fermion DM candidate. We constrain the model from the requirements of $(g-2)$, neutrino mass, LFV constraints and then discuss the singlet-doublet DM phenomenology. In Fig.~\ref{finalparam}, we showcase the final parameter space satisfying flavour observables as well as the constraints from correct relic density and direct search of DM in the plane of $m_{\chi_{3}}-\Delta M$. It is a riveting feature of this scenario that once the constraints from $(g-2)$ of electron and CLFV are imposed, it limits the allowed DM mass in a range $1-300$ GeV which gets further squeezed to around $60-300$ GeV once the LEP bound on $\psi^{\pm}$ mass is imposed ruling out the cyan coloured triangular region. It is also interesting to note that the mass splitting gets restricted only upto 20 GeV except in the Higgs resonance region where larger $\Delta M$ is allowed. This depicts the fact that in this scenario both dark sector phenomenology and the flavour observables are deeply coupled making it highly prognostic.  
	
	Thus, being in agreement with all relevant bounds, the model remains predictive at CLFV, DM direct detection, indirect detection as well as collider. In addition to the singlet-doublet parameter space sensitive to both high and low energy experiments like the LHC, MEG (or $(g-2)$) respectively, the existence of light $L_{\mu}-L_{\tau}$ at sub-GeV scale also remains sensitive at low energy experiments like NA62 at CERN, offering a variety of complementary probes.


	\noindent
	\acknowledgements
	DB acknowledges the support from Early Career Research Award from Science and Engineering Research Board (SERB), Department of Science and Technology (DST), Government of India (reference number: ECR/2017/001873). MD acknowledges DST, Government of India for providing the financial assistance for the research under the grant 
	DST/INSPIRE/03/ 2017/000032.
	
	\appendix 
	\section{Neutral Fermion Mass Matrix}
	\label{appendix1}
	Neutral fermion mass matrix for the dark sector in the basis $ ((\psi^0_R)^c, \psi^0_L, (N_{e})^c,(N_{\mu})^c,(N_{\tau})^c)^T$ as :
	\begin{eqnarray}
		\mathcal{M}&=&
		\left(
		\begin{array}{ccccc}
			0 &M & \frac{Y_\psi v}{\sqrt{2}}&0&0\\
			M &0 &\frac{Y_\psi v}{\sqrt{2}}&0&0\\
			\frac{Y_\psi v}{\sqrt{2}} &\frac{Y_\psi v}{\sqrt{2}} &M_{ee} &\frac{Y_{e\mu} v_ 1}{\sqrt{2}} & \frac{Y_{e\tau} v_ 1}{\sqrt{2}}\\
			0&0&\frac{Y_{e\mu} v_ 1}{\sqrt{2}}&  \frac{Y_{\mu}v_2}{\sqrt{2}}& M_{\mu \tau} \\
			0&0&\frac{Y_{e\tau} v_ 1}{\sqrt{2}}    &  M_{\mu \tau}    &
			\frac{Y_{\tau} v_2}{\sqrt{2}} \\
		\end{array}
		\right)\,\\
		&=&\left(
		\begin{array}{cc}
			\boldsymbol{M}&\boldsymbol{M_D} \\
			\boldsymbol{M^T_D}&\boldsymbol{M_R}
		\end{array}
		\right)
	\end{eqnarray}
	Where $\boldsymbol{M}=\left(\begin{array}{cc}
		0&M \\
		M&0
	\end{array}\right)$, $\boldsymbol{M_D}=\left(\begin{array}{ccc}
		\frac{Y_\psi v}{\sqrt{2}}&0&0\\
		\frac{Y_\psi v}{\sqrt{2}}&0&0
	\end{array}\right)$ and $\boldsymbol{M_R}=\begin{pmatrix}
		M_{ee} &\frac{Y_{e\mu} v_1}{\sqrt{2}} & \frac{Y_{e\tau} v_1}{\sqrt{2}}\\
		\frac{Y_{e\mu} v_1}{\sqrt{2}}&  \frac{Y_{\mu}v_2}{\sqrt{2}}& M_{\mu \tau} \\
		\frac{Y_{e\tau} v_1}{\sqrt{2}}    &  M_{\mu \tau}    &
		\frac{Y_{\tau} v_2}{\sqrt{2}} 
	\end{pmatrix}$.
	
	Since $\psi_L$ and $\psi_R$ has no coupling with $N_\mu$ and $N_\tau$ and $\boldsymbol{M_R}$ being a symmetric matrix can always be diagonalised using an orthogonal matrix $\mathcal{R}(\alpha_{12},\alpha_{13},\alpha_{23})$ such that the flavour eigen states($N_e,N_\mu,N_\tau$) are related to the mass eigenstates $N_1,N_2$ and $N_3$ (with masses $M'_1,M'_2$ and $M'_3$) as:
	\begin{eqnarray}
		N_e &=& c_{12} c_{13} N_1 +  (-c_{23} s_{12}-c_{12} s_{13} s_{23})N_2\nonumber\\&+& (-c_{12} c_{23} s_{13}+s_{12} s_{23})N_3\nonumber\\
		N_\mu &=& s_{12} c_{13} N_1 +  (c_{12} c_{23} -s_{12} s_{23} s_{13} )N_2\nonumber\\&+& (-s_{12} c_{23} s_{13}+c_{12} s_{23})N_3\nonumber\\
		N_\tau&=& s_{13} N_1+ c_{13} s_{23} N_2 + c_{13} c_{23} N_3
	\end{eqnarray}
	where we abbreviated $\cos \alpha_{ij} = c_{ij}$ and $\sin \alpha_{ij} = s_{ij}$.	
	
	As $\alpha_{ij}$ angles are free parameters, assuming $\sin \alpha_{12}$ and $\sin \alpha_{13}$ small, $N_e$ dominantly becomes $N_1$ with negligible admixture of $N_2$ and $N_3$.  
	
	Thus the neutral fermion mass matrix relevant for singlet-doublet DM phenomenology can be written in the basis  $ ((\psi^0_R)^c, \psi^0_L, (N_{1})^c)^T$ as :
	\begin{eqnarray}\label{sd_mass}
		\mathcal{M} & = &
		\left(
		\begin{array}{ccc}
			0 &M &c_{12}c_{13}\frac{Y_\psi v}{\sqrt{2}}\\
			M &0 &c_{12}c_{13}\frac{Y_\psi v}{\sqrt{2}}\\
			c_{12}c_{13}\frac{Y_\psi v}{\sqrt{2}} &c_{12}c_{13}\frac{Y_\psi v}{\sqrt{2}} &c^2_{12}c^2_{13}M'_{1}\\
		\end{array}
		\right)\,.\nonumber\\
		&=&\left(
		\begin{array}{ccc}
			0 &M &m_D\\
			M &0 &m_D\\
			m_D &m_D &M_1\\
		\end{array}
		\right)\,.
	\end{eqnarray}	
	
	Where $M_1= c^2_{12}c^2_{13}M'_{1}$ and $m_D=c_{12}c_{13}\frac{Y_\psi v}{\sqrt{2}}=c_{12}c_{13}m'_D$	
	
	\section{DM-SM Interaction}
	\label{appendix2}
	The interaction terms of the dark and visible sector particles in the gauged $U(1)_{L_\mu - L_\tau}$ scenario can be obtained by expanding the kinetic terms of $\Psi$ and $N_{R_i}$ given in Eq.-\eqref{sd_lag} as the following,
	
	\begin{equation}
		\footnotesize{
			\begin{aligned}
				\mathcal{L}_{\rm int} &= \overline{\Psi}i\gamma^\mu[-i\frac{g}{2}\tau.W_\mu - ig'\frac{Y}{2}B_\mu ]\Psi \\
				&+ \overline{N_{R_i}}i\gamma^\mu(-i g_{\small{\mu \tau}} Y_{\mu \tau}(Z_{\mu \tau})_\mu) N_{R_i} 
				\\ 
				&= \Big(\frac{e}{2\sin\theta_W \cos\theta_W}\Big)\overline{\psi^0}\gamma^\mu Z_\mu \psi^0
				\\
				& +\frac{e}{\sqrt{2}\sin\theta_W}(\overline{\psi^0}\gamma^\mu W^+_\mu \psi^- + \psi^+\gamma^\mu W^-_\mu \psi^0) \\
				& - e ~\psi^+\gamma^\mu A_\mu \psi^-\\ 
				& -\Big(\frac{e \cos2\theta_W}{2\sin\theta_W \cos\theta_W}\Big) \psi^+\gamma^\mu Z_\mu \psi^- \\
				&+Y_\psi \Psi \Tilde{H}(N_e+N^c_e).
		\end{aligned}}
	\end{equation}
	where $g = \frac{e}{\sin\theta_W}$ and $g'= \frac{e}{\cos\theta_W}$ with $e$ being the electromagnetic coupling constant, $\theta_W$ being the Weinberg angle and $g_{\mu \tau}$ is the $U(1)_{L_\mu - L_\tau}$ coupling constant. 
	
\vspace{1cm}	
These interactions, when written in terms of the physical states become
	{\small \begin{equation}
		\label{dmint}
		\begin{aligned}
			&\mathcal{L}_{\rm int}\\& =\Big(\frac{e}{2\sin\theta_W \cos\theta_W}\Big)(-\cos\theta\overline{\chi_{_{1L}}}i\gamma^\mu Z_{\mu}\chi_{_{2L}}\\&-\sin\theta\overline{\chi_{_{2L}}}i\gamma^\mu Z_{\mu}\chi_{_{3L}} + h.c.)\\
			& +\frac{e}{\sqrt{2}\sin\theta_W}(\cos\theta \overline{\chi_{_1}}\gamma^\mu W^+_\mu \psi^-\\&+ \overline{\chi_{_2}}i\gamma^\mu W^+_\mu \psi^- -\sin\theta \overline{\chi_{_3}}\gamma^\mu W^+_\mu \psi^-) \\
			& +\frac{e}{\sqrt{2}\sin\theta_W}(\cos\theta \psi^+\gamma^\mu W^-_\mu \chi_{_1}\\&-\psi^+i\gamma^\mu W^-_\mu \chi_{_2}-\sin\theta\psi^+ \gamma^\mu W^-_\mu\chi_{_3})\\
			& - e ~\psi^+\gamma^\mu A_\mu \psi^- 
			 -(\frac{e\cos2\theta_W}{2\sin\theta_W \cos\theta_W})~ \psi^+\gamma^\mu Z_\mu \psi^-.\\
		\end{aligned}
	\end{equation}}
	
%
	
	

\end{document}